\documentclass[12pt,a4paper]{article}
\pdfoutput=1
\usepackage{jcappub}
\usepackage{bm}
\usepackage{bbm}
\usepackage{etoolbox}
\usepackage{mhchem}
\usepackage[english]{babel}
\usepackage{multirow}
\usepackage{url}
\usepackage{caption}
\usepackage{subcaption}
\usepackage[utf8]{inputenc}
\usepackage[dvipsnames]{xcolor}

\usepackage{graphicx}
\usepackage{amsmath,amssymb}
\usepackage{hyperref}
\usepackage{color}
\usepackage{fleqn}
\usepackage{dcolumn}
\usepackage{soul}

\def\beq{\begin{equation}}
\def\eeq{\end{equation}}
\def\beqa{\begin{eqnarray}}
\def\eeqa{\end{eqnarray}}
\newcommand{\nl}{\nonumber \\}

\newcommand{\gag}{g_{a\gamma}}
\newcommand{\p}{\vec{p}}
\newcommand{\x}{\vec{x}}

\makeatletter
\gdef\@fpheader{}
\makeatother

\title{New bounds on Axion-Like Particles in the Ultraviolet from Legacy Data}

\author{Elisa Todarello} 
\emailAdd{elisa.todarello@berkeley.edu}

\affiliation{School of Physics and Astronomy, University of Nottingham,
University Park, NG7 2RD, Nottingham, United Kingdom}

\abstract{
We use legacy data from the Hubble Space Telescope (HST) and the International Ultraviolet Explorer (IUE) to search for a spectral line from the spontaneous decay of axion-like particle (ALP) dark matter. The HST data consist of blank sky observations taken with the Faint Object Spectrograph in the 165--240~nm wavelength range, while the IUE data consist of observations of the Virgo Cluster obtained with the long- and short-wavelength spectrographs, covering 195--325~nm and 123--200~nm, respectively.  
We set a 95\% C.L. upper limit on the ALP--photon coupling $g_{a\gamma} \lesssim 10^{-11}~\mathrm{GeV}^{-1}$ across the whole probed ALP mass range. Notably, we rule out values of $g_{a\gamma}$ above $2.3 \times 10^{-12}~\mathrm{GeV}^{-1}$ for ALP masses between 12.4 and 14.5\,eV, improving upon previous limits by a factor of seven.
}

\begin{document}
\maketitle

\section{\label{sec:intro} Introduction}
Axion-like particles (ALPs) are a compelling candidate for dark matter (DM). They are light pseudo-scalars that can be produced non-thermally in the early Universe~\cite{Arias:2012az}, similar to QCD axions~\cite{Peccei:1977hh, Peccei:1977ur, Weinberg:1977ma, Wilczek:1977pj, Dine:1982ah, Abbott:1982af, Preskill:1982cy}. Both QCD axions and ALPs couple to the electromagnetic field through the interaction term 
\[
\mathcal{L} = -\frac{1}{4}\gag\,a\,F_{\mu\nu}\tilde{F}_{\mu\nu}\enspace,
\]
where $a$ is the ALP field, $F_{\mu\nu}$ is the electromagnetic field strength tensor, $\tilde{F}_{\mu\nu} = \epsilon^{\mu\nu\rho\sigma}F_{\rho\sigma} / 2$ is its dual, and $\gag$ is the ALP--photon coupling constant. Through this interaction, an ALP can decay into two photons. If ALPs constitute the dark matter, their decay would produce a spectral line with a frequency equal to half the ALP mass, and a width determined by the DM momentum dispersion. The relative width of this line is of order $10^{-3}$ for dark matter in the Milky Way (MW) and can be broader in galaxy clusters or narrower in dwarf galaxies.

In this work, we use legacy data from the Hubble Space Telescope (HST) Faint Object Spectrograph (FOS)~\cite{Bahcall:1991} and from the International Ultraviolet Explorer (IUE)~\cite{Boggess:1978} to search for such a spectral line arising from the MW and Virgo cluster halos, respectively.
The HST FOS was operational from 1990 to 1997 and provided ultraviolet and optical spectra with high spatial and spectral resolution. We use data collected in 1991–1992 from blank sky regions, acquired during the science verification phase with the goal of studying the diffuse sky background~\cite{Lyons1993}. The IUE, which operated from 1978 to 1996, was one of the earliest space telescopes dedicated to ultraviolet astronomy. We use Virgo cluster observations taken with its short- and long-wavelength spectrographs in 1978, 1981, and 1988. Despite the age of these instruments, their legacy data continue to be valuable for searching for weak spectral lines from processes such as ALP decay.

The spectral line search technique employed here has been used in other works to set bounds on $\gag$ using spectroscopic data in the infrared~\cite{Yin:2023uwf, Janish:2023kvi, Pinetti:2025owq, Saha:2025any}, optical~\cite{Grin:2006aw, Todarello:2023hdk, Wang:2023imi}, and ultraviolet~\cite{Todarello:2024qci} bands. The ALP mass range between 12.4 and 14.5\,eV has remained relatively unexplored, with the strongest existing bound being $1.6 \times 10^{-11}~\mathrm{GeV}^{-1}$, derived from HST measurements of anisotropies in the cosmic optical background~\cite{Nakayama:2022jza, Carenza:2023qxh}. In this work, we significantly improve upon that limit \st{by one order of magnitude}, obtaining a bound $\gag < 4.6 \times 10^{-12}~\mathrm{GeV}^{-1}$ using the HST FOS, and $\gag < 2.3 \times 10^{-12}~\mathrm{GeV}^{-1}$ using the IUE short-wavelength spectrograph. 

This paper is structured as follows. In Section~\ref{sec:axion}, we derive expressions for the ALP decay signal, explicitly accounting for the ALP momentum distribution. In Section~\ref{sec:data}, we describe the data and the modeling of the expected signal for HST and IUE. Section~\ref{sec:res} presents our results. In Section~\ref{sec:compare}, we compare our findings to those from Ref.~\cite{Kar:2025ykb}. Finally, we conclude in Section~\ref{sec:conc}.

\section{\label{sec:axion} ALP signal}

We start by deriving an expression for the ALP spontaneous decay signal from the Boltzmann equation. We take explicitly into account the ALP momentum distribution.  
A more complex version of this calculation is presented in Refs~\cite{Caputo:2018vmy, Todarello:2023xuf} for the case of stimulated decay of ALPs into radio-photons. 
The rate of change of the ALP number density $n_a$ due to decays for which one of the photons has momentum $\vec{k}$ is given by
\beqa
\frac{d^3\dot{n}_a}{dk^3}
 &=& -
\frac{1}{(2\pi)^3}\frac{1}{2\omega_k}
\int \frac{d^3p}{(2\pi)^3}\frac{f_a(\vec{p}\,)}{2\omega_p}\ 
\frac{d^3q}{(2\pi)^3}\frac{1}{2\omega_q}
(2\pi)^4\delta^{(4)}(q + k - p)|\mathcal{M}_0|^2
\enspace.\label{boltzmann}
\eeqa
Here $p$ is the ALP 4-momentum, with $p^0=\omega_p = \sqrt{m_a^2 + {\vec{p}\,}^2}$, where $m_a$ is the ALP mass. The 4-momenta of the two photons produced in the decay are $k$ and $q$, with $k^0=\omega_k = |\vec{k}\,|$ and $q^0=\omega_q = |\vec{q}\,|$. The matrix element squared $|\mathcal{M}_0|^2 = \gag^2m_a^4/4$ is
summed over the polarizations of the two photons and includes a factor of 1/2 for identical particles in the final state.

We assume that the ALP phase-space distribution in the rest frame of the DM halo under consideration is at each point a Maxwell-Boltzmann, with an isotropic location-dependent width $\delta p(\x\,)$ and no net DM average momentum. The ALP DM phase space density is then
\beq
f_a(\p,\x\,) = n_a(\x\,) \frac{(2\pi)^3}{(2\pi)^{3/2}\delta p(\x\,)^3}\ e^{-\frac{{\p}^2}{2\delta p(\x\,)^2}}\enspace.\label{fa}
\eeq
The axion number density is given by
\beq
n_a(\x\,) = \int\frac{d^3p}{(2\pi)^3}\, f_a(\p\,,\x\,)\enspace. \label{na}
\eeq
Assuming ALPs are non-relativistic, we can neglect the ALP momentum everywhere in Eq.~\eqref{boltzmann}, except within the Dirac delta function and $f_a(\p\,)$. 
In fact, terms involving the ALP momentum are suppressed by a factor of $\sim 10^{-3}$ compared to those involving the ALP energy only. However, keeping track of the ALP momentum inside the Dirac delta leads to the $h_\nu$ factor below, which is relevant if we are observing the DM halo with an instrument that can resolve the ALP line width, set by $\delta p$.
After simple manipulations, we obtain the rate of change of the ALP number density per unit frequency and unit solid angle
\beqa
\frac{d\dot{n}_a}{d\nu d\Omega_k}
&=& -  \Gamma_a n_a \, h_\nu (\delta p) 
\qquad\qquad
h_\nu(\delta p) = \frac{1}{\sqrt{2\pi}\delta p}\, e^{-\frac{\epsilon^2}{2\delta p^2}}
\enspace,\label{dndot_dnudOmega}
\eeqa
where $\nu = \omega_k /(2\pi)$, $\epsilon = 4\pi\nu - m_a $, and $\Gamma_a = \gag^2m_a^3/(64\pi)$ is the ALP decay rate. Notice that $h_\nu (\delta p)$ reduces to a Dirac-delta $\delta(\epsilon)$ for $\delta p\to 0$, which is the relevant limit when the instrumental spectral resolution is much larger than the DM momentum dispersion.

Each of the two decay photons carries an energy approximately equal to $m_a/2$. Then, integrating along the line of sight $\ell$, we obtain the spectral intensity (or surface brightness) as
\beq
I_\nu = \int d\ell \ 2\, \frac{m_a}{2} \Big|\frac{d\dot{n}_a}{d\nu d\Omega_k}\Big|\, e^{-\tau_\nu}\enspace,
\eeq
where we have introduced the frequency-dependent optical depth $\tau_\nu$.  Integrating $I_\nu$ over the observing beam solid angle $B_\nu(\Omega)$, we obtain the flux density
\beq
S_\nu = \Gamma_a \int d\ell\int d\Omega\, B_\nu(\Omega)\ \rho_a(\x(\ell, \Omega))\, h_\nu\left(\delta p(\x(\ell, \Omega))\right) \, e^{-\tau_\nu(\x(\ell, \Omega))}\enspace,
\label{Snu}
\eeq
where $\rho_a = m_a n_a$ is the dark matter energy density.

We can now calculate the average flux density in a frequency bin of width $\Delta\nu$ centered at $\nu_0$. Defining $\epsilon_0 = 4\pi\nu_0 - m_a$, and assuming the beam and the optical depth don't vary appreciably over the bin width.  We obtain
\beqa
\langle S_\nu \rangle =
\frac{1}{\Delta \nu}\int_{\nu_0 - \frac{\Delta\nu}{2}}^{\nu_0 + \frac{\Delta\nu}{2}} d\nu\,S_\nu 
&=& \Gamma_a \int d\ell\,d\Omega\, B_{\nu_0}\, \rho_a\, \, e^{-\tau_{\nu_0}}\nl
&\times&\frac{1}{8\pi\Delta\nu}\left[\mathrm{erf}\left(\frac{\epsilon_ 0 + 2\pi\Delta\nu}{\sqrt{2}\delta p} \right)  -\mathrm{erf}\left(\frac{\epsilon_0 - 2\pi\Delta\nu}{\sqrt{2}\delta p} \right)   \right]
\enspace,\label{eq:Snu_generic_bw}
\eeqa
If the bin width is much larger than the dark matter momentum dispersion at any location, the expression above reduces to the familiar $\delta p$-independent form 
\beq
\langle S_\nu \rangle 
= \frac{\Gamma_a }{4\pi\Delta\nu\,} \Theta(2\pi\Delta\nu - |\epsilon_0|) \int d\ell\,d\Omega\, B_{\nu_0}\, \rho_a\, \, e^{-\tau_{\nu_0}}
\enspace,\label{eq:Snu_large_bw}
\eeq
where $\Theta$ is the Heaviside step function. We will use both the expressions above in obtaining our limits. A similar expression can be found for example in Ref.~\cite{Caputo:2018vmy} Eq.~(3.1), where a Bose-enhancement term $2f_\gamma$ is included and the Heaviside $\Theta$ is implied). After binning, the signal needs to be convolved with the detector energy response, called the line-spread function (LSF).

\section{\label{sec:data} Data and methods}

\renewcommand{\arraystretch}{1.3}
\begin{table}[]
    \centering
    \scriptsize
\begin{tabular}{|l|c|c|c|c|c|c|}
\hline
\multicolumn{1}{|c|}{obs. ids} & \# Files & $l$ [$^\circ$] & $b$ [$^\circ$] & $r_{min}$ [kpc] & $D$ [$10^{21}$eV/cm$^2$]  & Exp [s] \\
\hline \hline
y0g11802t &  1 & 113.22 & -66.76 &        8.12 &              1.832 &        150.00 \\ \hline
y0g11901t &  1 & 304.72 & -58.64 &        7.76 &              2.517 &        75.00 \\ \hline
y0g11a02t &  1 & 122.60 & -73.74 &        8.12 &              1.837 &        150.00 \\ \hline
y0g11c01t, y0g11c02t &  2 & 286.75 & -75.99 &        8.10 &              2.113 &        225.00 \\ \hline
y0g11f01t-y0g11f03t &  3 & 186.70 & -59.04 &        8.12 &              1.522 &        450.00 \\ \hline
y0g11l01t &  1 & 179.61 & -30.74 &        8.12 &              1.310 &        75.00 \\ \hline
y0g11m02t &  1 & 188.75 & -36.57 &        8.12 &              1.345 &        150.00 \\ \hline
y0g11p02t, y0g11p03t &  2 & 138.31 &  31.28 &        8.12 &              1.437 &        375.00 \\ \hline
y0g11q01t-y0g11q03t &  3 & 149.24 &  35.88 &        8.12 &              1.401 &        450.00 \\ \hline
y0g11r01t, y0g11r03t &  2 & 132.10 &  32.74 &        8.12 &              1.485 &        300.00 \\ \hline
y0g11s02t, y0g11s0bt &  2 & 129.61 &  31.50 &        8.12 &              1.499 &        300.00   \\ \hline
y0g11t03t &  1 & 207.74 &  45.72 &        8.12 &              1.450 &        225.00 \\ \hline
y0g11u01t-y0g11u03t &  3 & 171.04 &  49.88 &        8.12 &              1.438 &        450.00 \\ \hline
y0g11v03t &  1 & 238.71 &  40.54 &        8.12 &              1.610 &        225.00 \\ \hline
y0g11z03t &  1 & 192.59 &  53.21 &        8.12 &              1.472 &        225.00 \\ \hline
y0g12001t-y0g12003t &  3 & 232.14 &  50.78 &        8.12 &              1.616 &        450.00 \\ \hline
y0g12101t-y0g12103t &  3 & 180.63 &  62.67 &        8.12 &              1.560 &        435.00 \\ \hline
y0g12201t &  1 & 167.50 &  80.09 &        8.12 &              1.819 &        75.00 \\ \hline
y0g12401t, y0g12403t &  2 & 285.89 &  75.63 &        8.10 &              2.111 &        285.00 \\ \hline
y0g12501t, y0g12503t &  2 &  97.37 &  86.57 &        8.12 &              2.006 &        300.00 \\ \hline
y0g12602r &  1 & 120.19 &  46.45 &        8.12 &              1.650 &        150.00 \\ \hline
y0g12801t-y0g12803t &  3 & 148.17 &  78.80 &        8.12 &              1.822 &        450.00 \\ \hline
y0g12901t-y0g12903t &  3 & 117.11 &  45.26 &        8.12 &              1.672 &        225.00 \\ \hline
y0g12b03t &  1 & 110.99 &  53.17 &        8.12 &              1.771 &        225.00 \\ \hline
y0g12c01t, y0g12c02t &  2 &  18.61 &  68.68 &        7.62 &              2.627 &        225.00 \\ \hline
y0g12d02t &  1 & 337.83 &  48.84 &        6.44 &              3.519 &        150.00 \\ \hline
y0g12e01t-y0g12e03t &  3 & 351.74 &  56.93 &        6.84 &              3.220 &        450.00 \\ \hline
y0g12g01t, y0g12g03t &  2 &   7.36 &  41.14 &        5.40 &              4.381 &        300.00 \\ \hline
y0g12h01t, y0g12h03t &  2 &   7.23 &  41.02 &        5.39 &              4.394 &        300.00 \\ \hline
y0g12i01t, y0g12i03t &  2 &  94.76 &  42.04 &        8.12 &              1.939 &        300.00 \\ \hline
y0g12m01t-y0g12m03t &  3 &  57.97 &  -36.60 &       7.35 &              2.840 &        450.00\\ \hline
\end{tabular}

\caption{Summary of HST FOS data used in this work. The columns are: observation identifier, number of files for a given pointing of the telescope, galactic coordinates ($\ell$, $b$) of the pointing, distance of closest approach of the line of sight to the Galactic center $r_{min}$, D-factor obtained using the NFW of \texttt{MilkyWayPotential2022} from the package \texttt{gala}~\cite{2017JOSS....2..388P}, and total exposure per wavelength bin (also called pixel).}
\label{tab:data_FOS}
\end{table}

\subsection{Hubble Space Telescope (HST) Faint Object Spectrograph (FOS)}

\begin{figure}[t!]
\centering
   \includegraphics[width=0.9\textwidth]{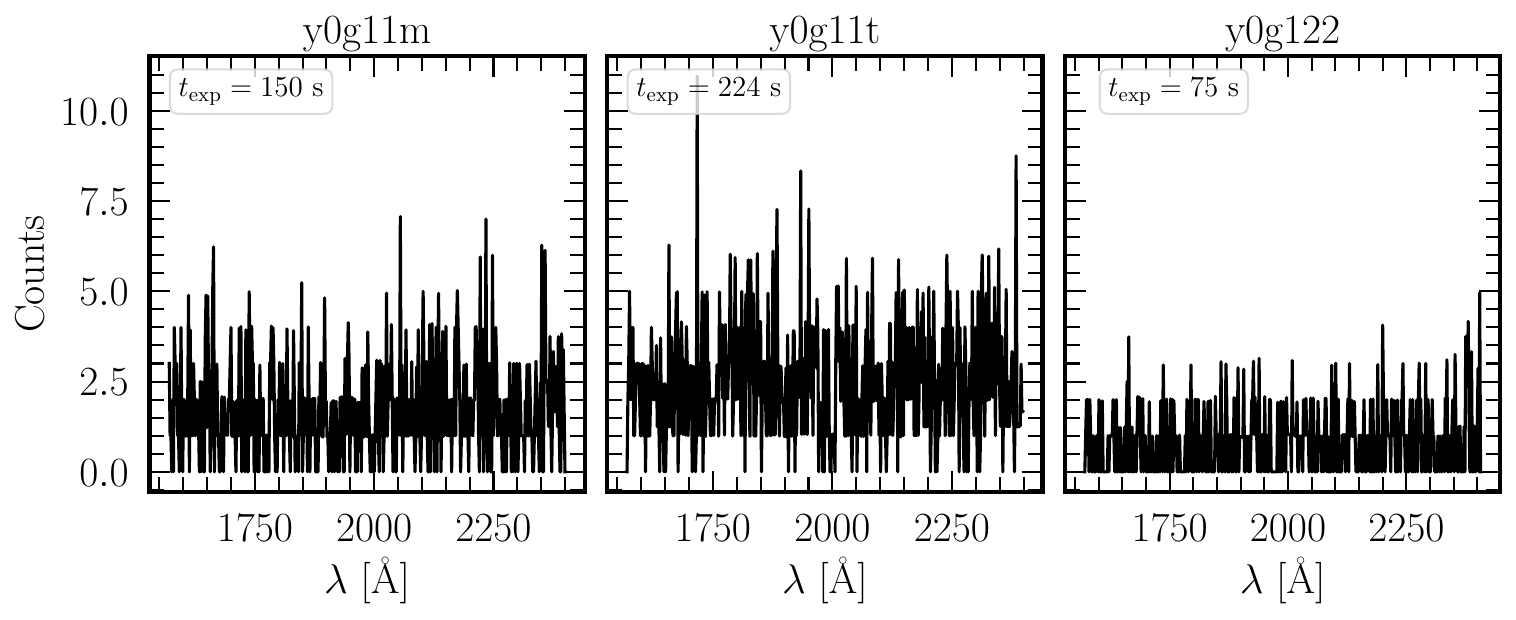}
    \caption{Three of the stacked HST FOS spectra used in this work. Each pointing is identified by the first six characters of the filenames that entered the stack.
    }
\label{fig:spectra_FOS}
 \end{figure}

\subsubsection{Data}\label{sec:data_HST}
We analyze blank sky observations acquired with the Hubble Space Telescope (HST) using the Faint Object Spectrograph (FOS) from February 1991 to March 1992, during the science verification phase~\cite{Lyons1993}.
The goal of these observations was to study the sky background. Data were acquired under three proposal numbers: SV2965 (high galactic latitude), SV2966 (low galactic latitude),
and SV2967 (low ecliptic latitude).
Here we will use the high galactic latitude data set. 

These data were acquired before the installation of the Corrective Optics Space Telescope Axial Replacement (COSTAR). COSTAR was introduced to correct the spherical aberration caused by a flaw in the HST primary mirror, affecting the FOS~\cite{FOSHandbook1998}. The aberration produced extended outer wings of the point spread function and consequent source acquisition failures, particularly in crowded fields, until COSTAR was deployed in 1994. However, the observations we use here targeted the diffuse sky background, and should therefore not be significantly affected by these aberration effects.

The FOS had two Digicon detectors: the blue detector (1150-5400~\AA) and the red or amber detector (1620-8500~\AA). The Digicons functioned by accelerating photoelectrons produced by a photocathode onto a linear array of 512 silicon diodes. The red detector was more sensitive than the blue one in the overlap region~\cite{FOSHandbook1998}. Additionally, we find that data from the blue detector are heavily contaminated by spurious spectral lines. For these reasons, our analysis focuses exclusively on the red-detector data. 
Moreover, we only use observations taken with the G160L grating, which is sensitive in the wavelength range of interest.

This leaves us with 113 data files available on MAST~\cite{mast} with 37 different pointings toward given galactic coordinates. 
For all but a couple of lines of sight, there are three data files. The first file contains two exposures for a total of 300 seconds, the second, 5 exposures for a total of 600 seconds, and the third, 7 exposures for a total of 900 seconds. The exposure per wavelength bin (also called pixel) is given by the total exposure divided by 4, because the spectra were shifted in one-quarter diode increments, resulting in 2064 wavelength bins~\cite{FOSHandbook1998}. The lowest galactic latitude for observations in this set is $|\ell|=30.7^\circ$ and the lowest distance of closest approach of any line of sight to the Galactic center is 5.39~kpc. 

The target acquisition aperture (\texttt{A-1} or \texttt{4.3}) was used for all data. This aperture was a $4.3''\times 4.3''$ square (pre-COSTAR). Since the height of a diode was $1.43''$, the effective collecting area was $4.3'' \times 1.43''$.

The spectral resolution is not a strong function of wavelength and can be derived from Tables 29.4 and 29.5 of~\cite{FOSHandbook1998}. With the G160L grating and \texttt{4.3} aperture, used for our data, we obtain a full width at half maximum (FWHM) spectral resolution of~81~\AA~(48 pixels) for extended sources filling the aperture, like the signal from ALP decay. No data taken with high-dispersion gratings is available within SV2965.

We analyze spectra in units of photon counts, which can easily be derived from intermediate products of the FOS calibration pipeline.  Details of the FOS calibration pipeline and creation of the count spectra can be found in Appendix~\ref{app:data_reduction}.

We apply the following data cut. For each of the 113 observations, we calculate the integrated count rate $I$ and discard observations that have above-median $I > 0.007$~counts/s/\AA.
The data files that survive the cut are listed in Table~\ref{tab:data_FOS}. We then stack spectra sharing the same line of sight, ending up with 31 data files for a total exposure of 8595 seconds per pixel\footnote{The stacked spectra are available for download at \href{https://github.com/elisabm99/UV-legacy}{UV-legacy repository}.}.
Three example spectra are shown in Figure~\ref{fig:spectra_FOS}.

\subsubsection{Modeling of the signal}\label{sec:model_HST}
In order to bracket the uncertainty on our bound due to the dark matter energy density of the Milky Way, we consider several profiles.
For our analysis, we take the Navarro-Frenk-White (NFW) profile~\cite{Navarro:1995iw}
\beq
\rho_{\rm{NFW}}(r)=\frac{\rho_s}{\left(\frac{r}{r_s}\right)\left( 1 + \frac{r}{r_s} \right)^2}\enspace,
\label{eq:rho}
\eeq
from \texttt{MilkyWayPotential2022} from the galactic dynamics package \texttt{gala}~\cite{2017JOSS....2..388P}, with parameters $\rho_s=0.439$~GeV~cm$^{-3}$ and $r_s=15.62$~kpc.
In Eq.~\eqref{eq:rho}, $\rho_s$ and $r_s$ are the scale density and radius, respectively. This profile fits well various measurements of the MW enclosed mass up to galactic radii $>100$~kpc, as well as measurements of the MW rotation curve between 5 and 20~kpc~\cite{2025NewAR.10001721H}.
Notice that our results are completely insensitive to the structure of the Galactic DM halo within 5.39~kpc from its center.
We further consider two profiles from~\cite{2020MNRAS.494.4291C}. Both models have been fitted to the Gaia rotation curve from~\cite{2019ApJ...871..120E} and the total MW mass from ~\cite{2019MNRAS.484.5453C}.
The first is best-fit NFW from~\cite{2020MNRAS.494.4291C} with parameters $\rho_s=0.438$~GeV~cm$^{-3}$ and $r_s=14.84$~kpc. The second is their best-fit contracted profile, derived by starting with an NFW with $\rho_s=0.184$~GeV~cm$^{-3}$ and $r_s=22.3$~kpc and applying a contraction due to the gravitational pull of baryons in the Galaxy located at the center of the halo. The functional shape of the contraction is informed by the structure formation simulations Auriga~\cite{Grand:2017rdp}, APOSTLE~\cite{Fattahi:2016nld, Sawala:2016tjm}  and EAGLE~\cite{Schaye:2014tpa}.

In order to allow comparison with works similar to ours, we consider three further NFW profiles. We will denote the results from these profiles as ``extended analysis". The first has $\rho_s=0.256$~GeV~cm$^{-3}$ and $r_s=19.1$~kpc. It is taken from the 2D 68\% containment region for $M^{\text{DM}}_{200}$ and $c^{\text{NFW}}$ of~\cite{2020MNRAS.494.4291C}
and was used in~\cite{Roy:2023omw}. The second 
has $\rho_s=0.184$~GeV~cm$^{-3}$ and $r_s=24.42$~kpc. It is from~\cite{Cirelli:2010xx} and
was used in~\cite{Pinetti:2025owq, Saha:2025any}. The third and final profile, taken from~\cite{Cirelli:2024ssz}, was used in~\cite{Regis:2024znx} and has $\rho_s=0.566$~GeV~cm$^{-3}$ and $r_s=14.46$~kpc.

Finally, we consider the Einasto profile from~\cite{2024MNRAS.528..693O}, parametrized as
\beq
\rho_{\rm{Einasto}}(r)=\rho_s\, e^{-\left( \frac{r}{r_s}\right)^\alpha}\enspace,\label{eq:rho_einasto}
\eeq
with $\rho_s = 3.26$~GeV~cm$^{-3}$, $r_s=3.86$~kpc and $\alpha=0.91$. While this profile is comparable to the NFW ones in the range $r=5-15$~kpc, it drops more rapidly at larger radii and may, in principle, yield different bounds.

Throughout, we use a distance of the Sun from the Galactic center of $R = 8.122$~kpc~\cite{2018A&A...615L..15G} and the Galactic extinction law from Ref.~\cite{1989ApJ...345..245C} with magnitude along the line of sight from~\cite{extinct}. Since the spectral resolution is much larger than the signal's intrinsic bandwidth, we compute the signal using Eq.~\eqref{eq:Snu_large_bw}. We then convolve it with the instrument extended-source LSF, which we model as a Gaussian with a FWHM of 81~\AA.

We relate detector counts to the physical signal by constructing a forward model based on the FOS calibration pipeline reference files. The expected photon count at the detector is
\beq
C(\lambda) = R(\lambda) S_\lambda \enspace,\label{eq:counts}
\eeq
where $S_\lambda$ is the ALP-decay flux entering the aperture, per unit wavelength, convolved with the LSF. The forward model operator, $R(\lambda)$, accounts for the instrument's effective sensitivity and is defined as
\beq
R(\lambda) = \frac{T(\lambda)}{\rm IVS_{e}(\lambda)}\, t_{\rm exp}\enspace,\label{eq:forward}
\eeq
where $T(\lambda)$ is the time-dependent sensitivity correction,  ${\rm IVS_{e}(\lambda)}$ is the extended-source inverse sensitivity curve, as described in detail in Appendix~\ref{app:data_reduction}, and $t_{\rm exp}$ is the exposure time per pixel.

\subsection{International Ultraviolet Explorer (IUE)}\label{sec:IUE}
\subsubsection{Data}

\begin{figure}[t!]
\centering
   \includegraphics[width=0.9\textwidth]{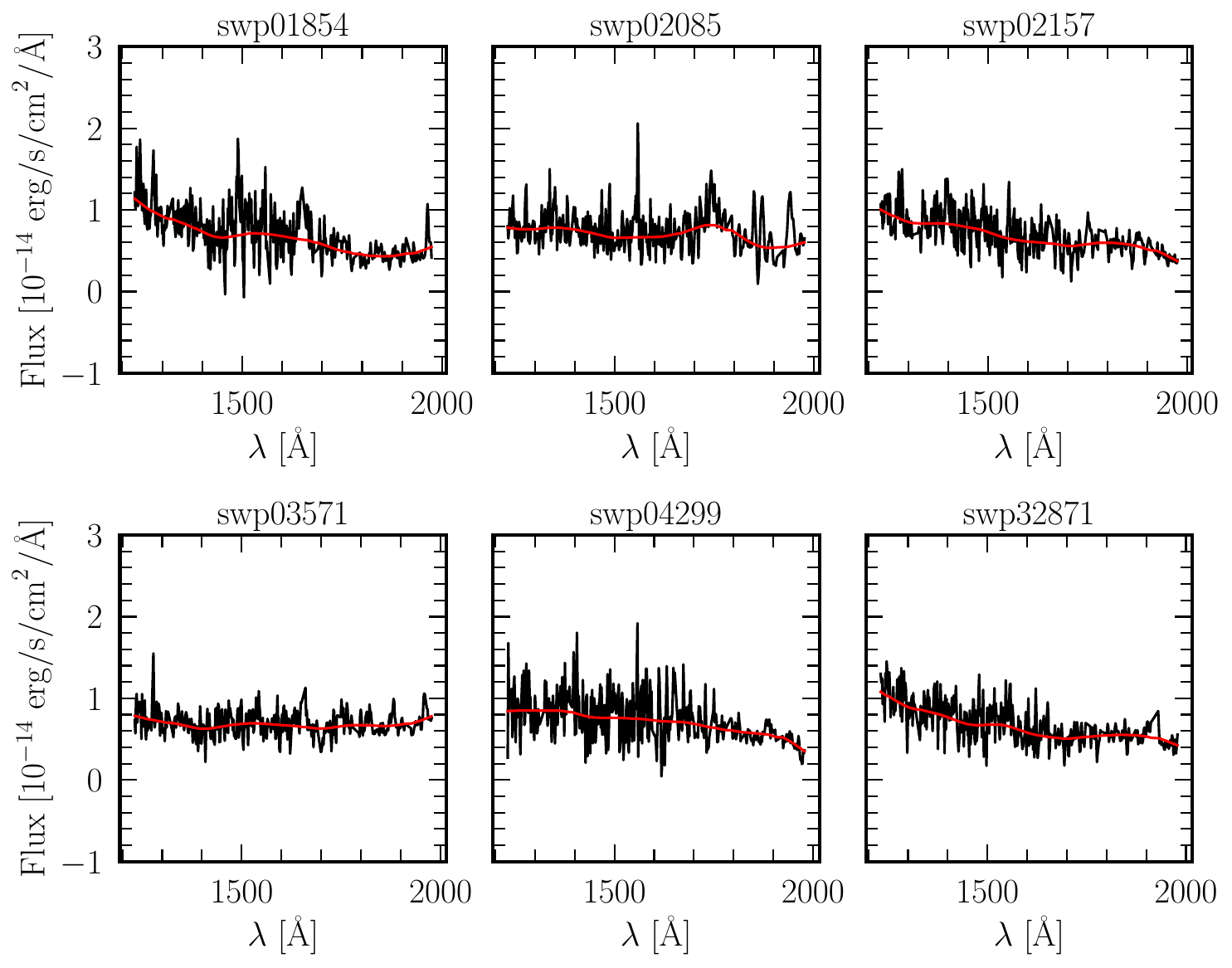}
    \caption{IUE spectra acquired with the short-wavelength spectrograph. Red lines mark the continuum model.
    }
\label{fig:spectra_sw}
 \end{figure}

\begin{figure}[t!]
\centering
   \includegraphics[width=1\textwidth]{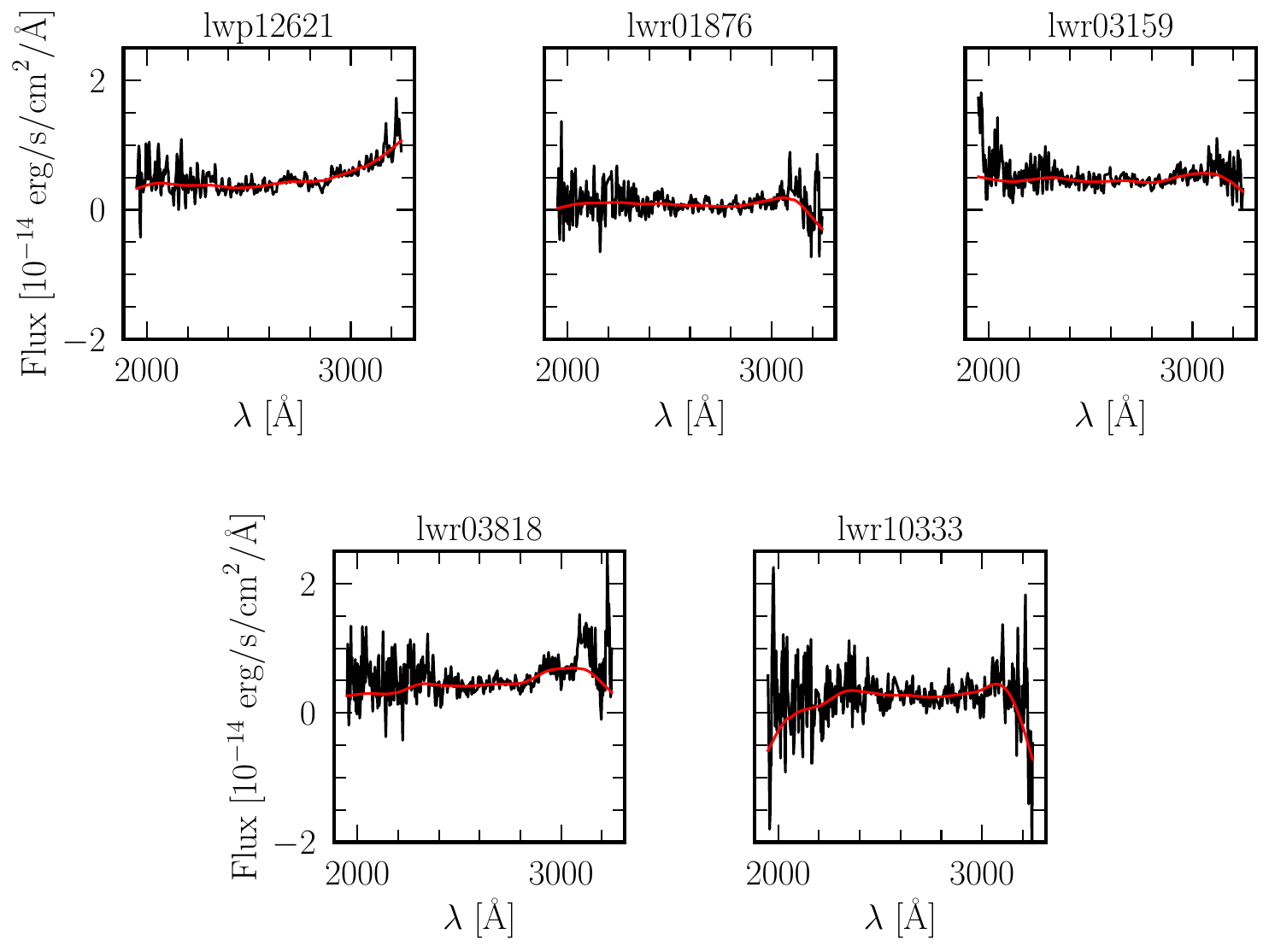}
    \caption{IUE spectra acquired with the long-wavelength spectrograph. Red lines mark the continuum model.
    }
\label{fig:spectra_lw}
 \end{figure}
 
\renewcommand{\arraystretch}{1.3}
\begin{table}
\scriptsize
\centering
\begin{tabular}{|c|c|c|c|c|}
    \hline 
    obs. id   & exposure [ks] & ang. sep. 
    [$'$] & ang. sep. M87 [$'$] & $\tilde{D}$ [$10^{24}$~eV/cm$^2$] \\ \hline \hline 
    lwp12621  & 21.18 & 0.553 & 0.079 & 1.51 \\ \hline 
    lwr01876  & 20.40 & 0.510 & 0.165 & 1.49 \\ \hline 
    lwr03159  & 23.40 & 0.510 & 0.165 & 1.49 \\ \hline 
    lwr03818  & 13.80 & 0.416 & 0.144 & 1.57 \\ \hline 
    lwr10333  & 3.60  & 0.542 & 0.087 & 1.46 \\ \hline 
    \hline 
    swp01854  & 18.00 & 0.564 & 0.037 & 1.48 \\ \hline 
    swp02085  & 22.98 & 0.510 & 0.165 & 1.53 \\ \hline 
    swp02157  & 24.00 & 0.601 & 0.109 & 1.46 \\ \hline 
    swp03571  & 22.20 & 0.510 & 0.165 & 1.53 \\ \hline 
    swp04299  & 25.80 & 0.416 & 0.144 & 1.62 \\ \hline 
    swp32871  & 44.88 & 0.495 & 0.156 & 1.55 \\ \hline 
\end{tabular}

\caption{IUE observations used in this work. The columns are: the observation identifier, the total exposure, the angular separation from our reference center and from the center of M87, and the modified D-factor $\tilde{D}$ defined in Eq.~\eqref{D_tilde}, calculated using the parameters of our fiducial analysis. }
\label{tab:data_IUE}
\end{table}

We analyze all International Ultraviolet Explorer (IUE) data with pointing within one arcminute of M87 available on the IUE Newly Extracted Spectra (INES) system~\cite{Rodriguez-Pascual:1999bwf}.
We find six spectra collected with the short-wavelength spectrograph (1150-2000~\AA), and five spectra collected with the long-wavelength spectrograph (1850-3300~\AA), one with the prime and four with the redundant camera. All considered data are taken in the low-dispersion mode and through the large apertures.

The INES system was created to address concerns~\cite{Rodriguez-Pascual:1999bwf, Massa:1998nr} regarding the calibration of low-dispersion spectra in the New Spectroscopic Image Processing System (NEWSIPS)~\cite{1996AJ....111..517N}. In particular, the INES system takes the two-dimensional, spatially resolved, rotated images (SILO files) created by NEWSIPS, and extracts one-dimensional spectra, applying improved noise models, background determination, and extraction profile. Two additional spectra with pointing within $1'$ of M87 are available on MAST~\cite{mast}. However, since they were calibrated with NEWSIPS and not with INES, we do not include them in our analysis.

The sizes of the large apertures as projected on the cameras are reported in~\cite{1997IUENN..57....1G} (Section 2.2). IUE large apertures are slots, with a size of approximately $10''\times 20''$. In the following, we will approximate them as circles with radius $R=\sqrt{A/\pi}$, where $A$ is the area given in~\cite{1997IUENN..57....1G}.  The impact of this approximation on our bounds is subdominant compared to other sources of systematic errors that we account for.
The projection of the aperture on the detectors spans approximately 14 pixels in the cross-dispersion direction~\cite{newsips1997}.

The INES spectral extraction algorithm is described in Refs.~\cite{Rodriguez-Pascual:1999bwf, RodriguezPascual1998ESASP413_731}. The
spectrum is extracted at each wavelength from the SILO 2D image, from a window centered on the main spectral trace, and 23-pixel wide in the cross-dispersion
direction. The background is
estimated from two 7-pixel swaths placed symmetrically on both sides of
the aperture region at a distance of 13 pixels from the spectral trace. The background measured in these
swaths is iteratively smoothed along the dispersion direction with a
31-pixel boxcar filter and outlier rejection, and then linearly interpolated
across the aperture window. 
The background swaths are far enough from the projection of the aperture on the detector that any contribution of a diffuse ALP signal is expected to be negligible.
After background subtraction, INES determines an empirical extraction profile \(p(x,\lambda)\) based on the background-subtracted image according to Eq~(1) of~\cite{Rodriguez-Pascual:1999bwf}. Here $x$ labels  pixels in the cross-dispersion direction. If the source is too faint to perform empirical extraction, INES defaults to a 23-pixel wide boxcar extraction profile, which is the case for two of our spectra (lwr01876 and lwr10333).

The spectral resolution of IUE's spectrographs is approximately 6~\AA. We take the FWHM resolution along the dispersion direction from Figure 2.19 of~\cite{1997IUENN..57....1G}, interpolating linearly between data points. 
In creating the SILO files, NEWSIPS applies a de-tilting of the 2D image if the source is classified as extended, i.e. filling the aperture~\cite{1996AJ....111..517N}. This adjustment is necessary because the large aperture's major axis is not aligned perpendicularly to the dispersion direction. Extracting the 1D spectrum of an extended source without de-tilting by simply summing the flux perpendicular to the dispersion results in a degradation of the spectral resolution of 20 to 25\%.
De-tilting is not necessary for point sources that are localized within the aperture.
The emission from the decay of ALP dark matter we search for here is an extended source. However, some of the spectra we consider were treated as point sources. For those, we multiply the FWHM spectral resolution from  Figure 2.19 of~\cite{1997IUENN..57....1G} by a factor 1.25.

The spectra are shown in Figures~\ref{fig:spectra_sw} and~\ref{fig:spectra_lw} and listed in Table~\ref{tab:data_IUE}, along with the exposure, angular separation from the center of the DM halo used for our fiducial analysis and from M87, and modified D-factor (see Section~\ref{sec:model_IUE}). We remove wavelength bins flagged for data quality as well as the region $\lambda < 1230$~\AA, to remove the bright Ly$\alpha$ emission. 
For long-wavelength spectrograph data, we restrict the range to 1950~\AA~$<\lambda<$~3250~\AA.
The red lines in Figures~\ref{fig:spectra_sw} and~\ref{fig:spectra_lw} mark the continuum model that we subtract from the data. For our search, further background subtraction is appropriate in addition to what is already done in INES.  While INES accounts for instrumental backgrounds such as radioactive decay, sky background, scattered light, and readout noise, in our case, continuum emission from the target of observation (M87) also constitutes a background.
To determine the continuum level, we use the asymmetrically reweighted Penalized Least Squares algorithm (arPLS)~\cite{2015Ana...140..250B} with a smoothness parameter of $10^5$. Given the narrow width of a potential ALP decay line, its presence would not significantly affect the arPLS continuum estimate.

\subsubsection{Modeling of the signal}\label{sec:model_IUE}
\begin{figure}[t!]
\centering
   \includegraphics[width=0.7\textwidth]{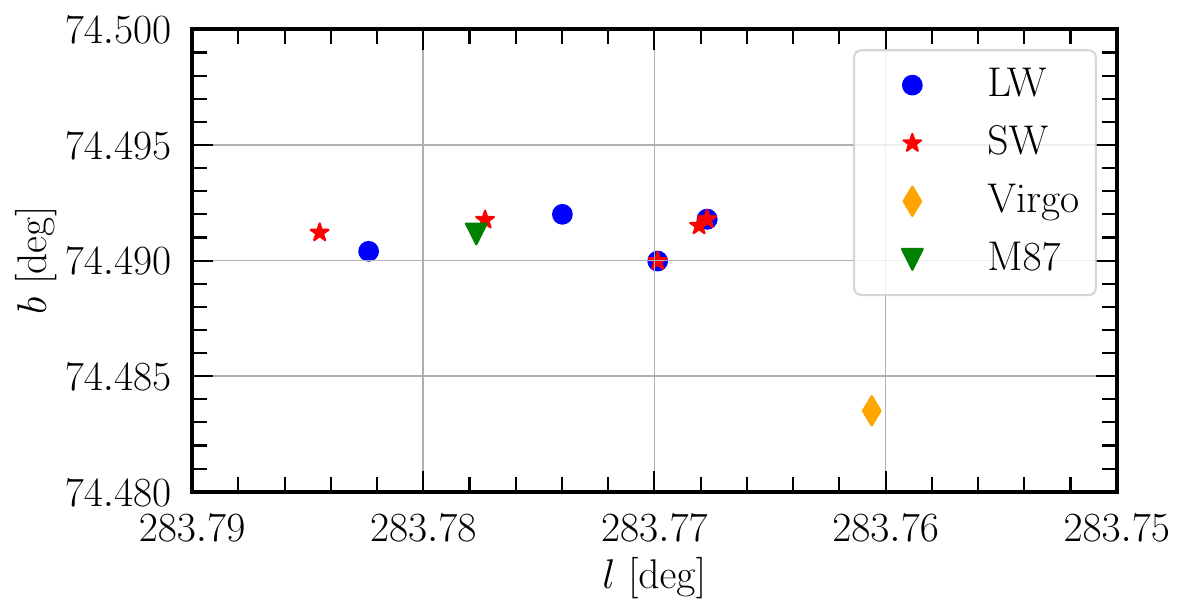}
    \caption{Galactic coordinates of the observations taken with the short (red stars) and long (blue dots) wavelength spectrographs. The green triangle marks the position of M87, while the orange diamond is the assumed position of the center of the Virgo DM halo for our fiducial analysis.
    }
\label{fig:IUE_coords}
 \end{figure}
For our fiducial analysis, we adopt the DM profile for the Virgo dark halo from~\cite{McLaughlin:1998sb}. It is an NFW profile with scale density $\rho_s =3.2\times10^5\,M_\odot$~kpc$^{-3}$ and scale radius $r_s=560$~kpc. 
We adopt a distance $d=15.9$~Mpc to the center of the Virgo cluster~\cite{Tully:2016ppz} and a redshift $z=0.0038$. 
Finally, we choose the reference center of the DM halo of Virgo from~\cite{Sazonov:2024fnz}, with galactic coordinates $l=283.7606^\circ$ and $b=+74.4835^\circ$. 

To estimate the systematic uncertainty on our bound due to modeling of the DM halo, we vary $\rho_s$ and $r_s$ within the error bars reported in~\cite{McLaughlin:1998sb}: $\rho_s =(3.2^{+2.6}_{-1.3})\times10^5\,M_\odot$~kpc$^{-3}$ and $r_s=560_{+200}^{-150}$~kpc. We also consider two additional estimates of the distance to the center of Virgo, $d = 16.5$~Mpc from~\cite{Mei:2007xs} and $d = 15.46$~Mpc from~\cite{DiMauro:2023qat}. Finally, we explore the uncertainty due to the position of the center of the DM halo. From Figure~\ref{fig:IUE_coords}, it's clear that assuming the center of the DM halo to coincide with M87, whose position we take from~\cite{2020A&A...644A.159C}, will yield a more stringent bound than our fiducial analysis since observations are focused there. The angular separation between M87 and our reference center is about $30''$, corresponding to a projected distance of about 2.5~kpc. To bracket the uncertainty on our bound, we will consider the case in which the center of the halo coincides with M87, and the case in which the center is $30''$ away from our reference center in the direction opposite to M87.

\begin{figure}[t!]
\centering
   \includegraphics[width=\textwidth]{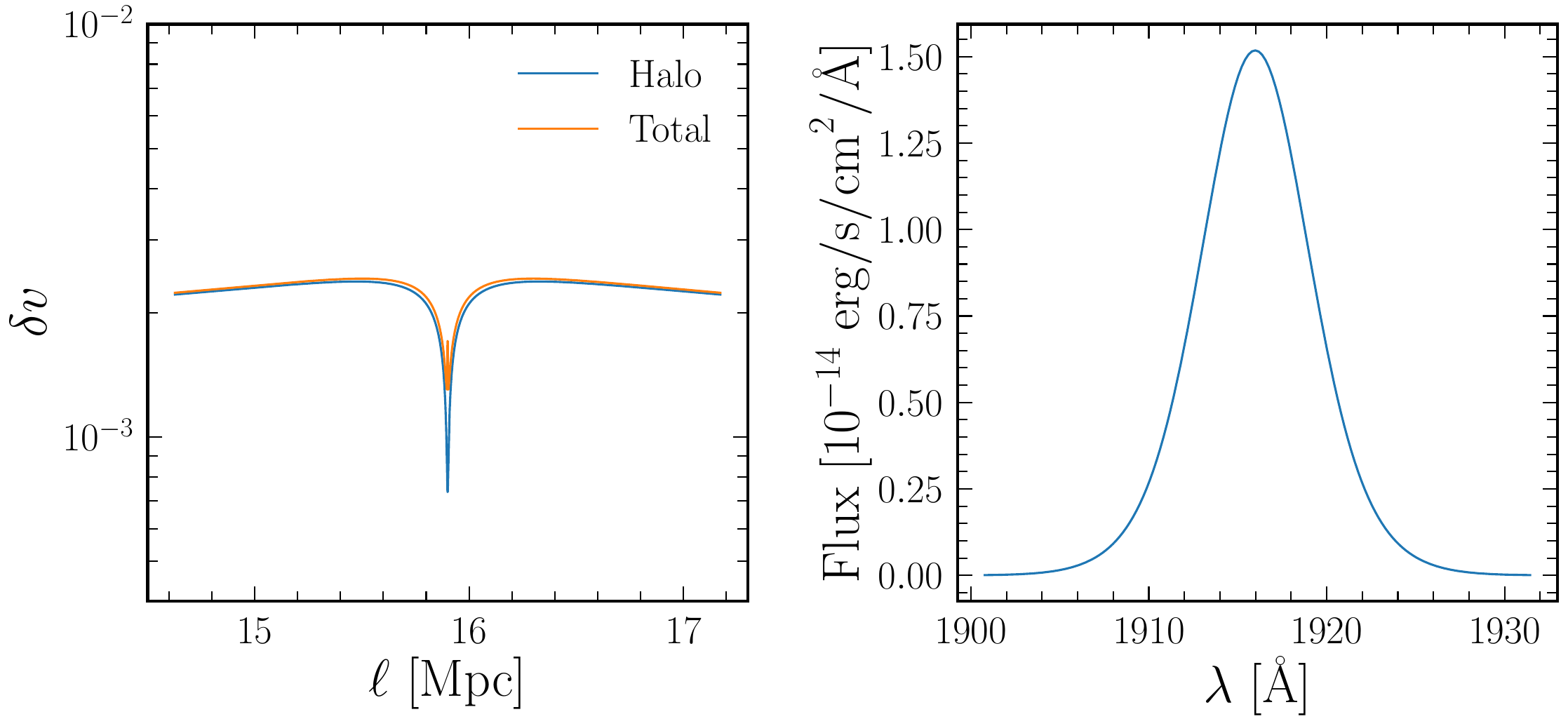}
    \caption{\textbf{Left}: Velocity dispersion as a function of the distance along the line of sight $\ell$, for a line of sight passing through the center of the swp01854 observation. The orange line shows the total velocity dispersion, given by the halo contribution (blue line) plus the BH contribution. 
    \textbf{Right}: Expected signal for observation swp01854, $m_a = 13$~eV and $\gag = 10^{-12}~\mathrm{GeV}^{-1}$, assuming the modeling of our fiducial analysis.
    }
\label{fig:dv_and_signal}
 \end{figure}

We compute the velocity dispersion $\delta v(r)$ integrating the spherically symmetric Jeans equation assuming no velocity anisotropy (see for example~Ref.~\cite{Lokas:2000mu}). We consider an additional component of the velocity dispersion around the supermassive black hole (BH) M87*, $\delta v_{BH} = \sqrt{2GM_{BH} / r}$, where $M_{BH} = 6.5\times 10^9~M_\odot$~\cite{EventHorizonTelescope:2024uoo} and $r$ is the distance from the BH position. We conservatively assume that the distance to the BH is the same as the distance to the center of the DM halo\footnote{This choice implies that the large velocity dispersion around the black hole suppresses the contribution of the NFW cusp to $\tilde{D}$.}. The left panel of Figure~\ref{fig:dv_and_signal} shows the velocity dispersion along the line of sight to the center of the swp01854 observation. The component due to the halo is consistent with measurements of the velocity dispersion~\cite{Girardi:1995iy, McLaughlin:1998sb} of order 640~km/s.

For each spectrum, Table~\ref{tab:data_IUE} reports the angular separation from the reference center of the DM halo and from M87, and modified D-factor
\beq
\tilde{D} = \langle\delta v\rangle\int_{0}^{\infty} d\ell\int d\Omega\,B(\Omega)\,\frac{ \rho_a}{\delta v}
\qquad\qquad
\langle\delta v\rangle = \frac{1}{V}\int_{d-r_s}^{d+r_s} d\ell\,\ell^2\int d\Omega\,B(\Omega)\,\delta v
\enspace,\label{D_tilde}
\eeq
where $V$ is the volume of the integration region. The D-factor due to the MW halo along the same line of sight is typically a factor of 20 smaller than that due to Virgo.

We apply the dust extinction in the Virgo intra-cluster medium from~\cite{virgodust}, with an extinction $A_V=0.14$~mag and a 
Magellanic Cloud-like extinction law~\cite{2003ApJ...594..279G}, i.e. with the functional shape of~\cite{1989ApJ...345..245C} and $R_V= 3.41$. Galactic extinction is modeled in the same way as for HST data. The expected signal for swp01854, with an ALP mass of 13~eV and coupling $\gag = 10^{-12}~\mathrm{GeV}^{-1}$, is shown on the right panel of Fig.~\ref{fig:dv_and_signal}. It has a FWHM of about 7.3~\AA, comparable to IUE's spectral resolution.
Then, we compute the flux density as in Eq.~\eqref{eq:Snu_generic_bw} and convolve it with the detector response, which we take to be a Gaussian with a standard deviation $\sigma_\lambda = \mathrm{FWHM} / (2 \sqrt{2 \log(2)})$ and FWHM from~\cite{1997IUENN..57....1G}.

To compare with the data, we closely reproduce the spectral extraction procedure of~\cite{Rodriguez-Pascual:1999bwf, RodriguezPascual1998ESASP413_731}
to reconstruct the extraction profile \(p(x,\lambda)\) and estimate how detector lines were
down-weighted in the published 1D spectra. We use the publicly available INES noise models~\cite{SchartelRodriguezPascual1998INESNoiseModel}. Under the simplifying
assumption that the ALP surface brightness is uniform across the
aperture, we define an effective number of contributing cross-dispersion
lines as $ N_{\rm eff}(\lambda) = 1/\sum_x p(x,\lambda)^2$, where $x$ labels the position in the 23-pixel-wide extraction window.
We normalize \(p\) such that \(\sum_x p(x,\lambda)=1\) at each wavelength. Whenever
\(N_{\rm eff}(\lambda)\) is smaller than the geometric number of lines
subtended by the aperture, \(N_{\rm geom}\), we down-weight the expected
signal by \(N_{\rm eff}(\lambda)/N_{\rm geom}\).

\section{Results}\label{sec:res}
\begin{figure}[t!]
\centering
   \includegraphics[width=0.9\textwidth]{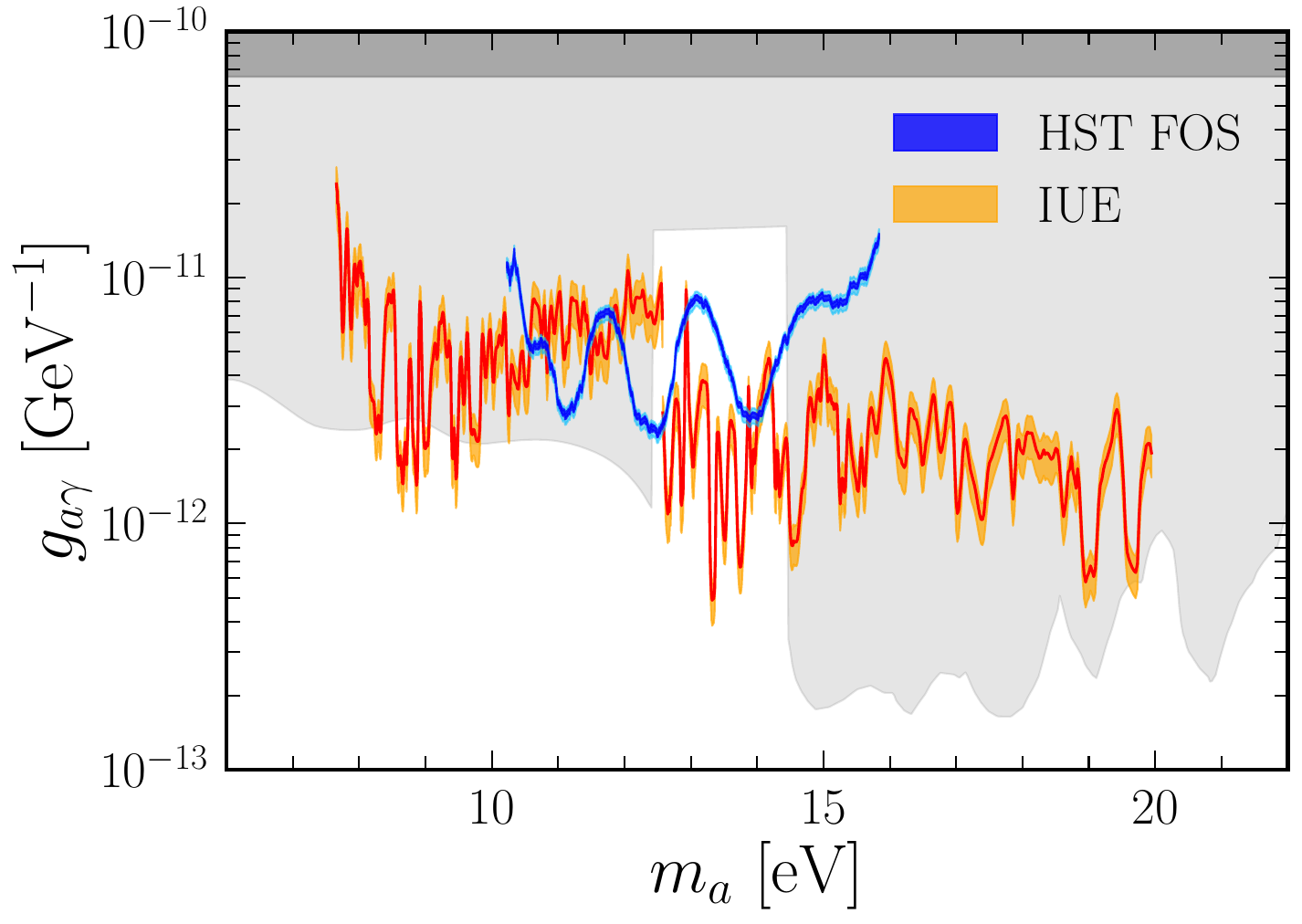}
    \caption{95\% C.L. upper bounds obtained from HST FOS (blue) and IUE (red/orange) data. The thickness of the colored bands accounts for systematic errors due to the modeling of the signal. The dark gray shaded regions show existing bounds from evolution in globular clusters~\cite{Ayala:2014pea}, while the region in light gray is excluded by Refs.~\cite{Carenza:2023qxh,Porras-Bedmar:2024uql,Wang:2023imi, Wadekar:2021qae, Bolliet:2020ofj, Todarello:2024qci}.
    }
\label{fig:result}
 \end{figure}

\subsection{HST FOS}\label{sec:res_FOS}
For each ALP mass, we fit the data to our model in a 250-pixel window centered at $\lambda_a = 4\pi/m_a$. This window is chosen to be much wider than the expected ALP-induced feature, which spans about $48$ pixels after convolution with the instrumental line-spread function, so that the fit can constrain a smooth background simultaneously with the signal template, while still being narrow enough that the background can be accurately approximated by a low-order polynomial. 
The expected photon count in each pixel $i$ of the stacked data set $j$ is given by
\beq
\lambda_{ij} = g_{10}^2 C_{ij} + b_{ij}\enspace,
\eeq
where $g_{10} = \gag / (10^{-10}~\mathrm{GeV}^{-1})$, $C_{ij}$ is the ALP photon count if $\gag = 10^{-10}~\mathrm{GeV}^{-1}$, and $b_{ij}$ is a smooth background model. This background term is meant to represent the detector's dark count plus any continuum sky emission that is smooth over the scale of the 48-pixel wide signal. We model this term as a second-order polynomial $b_j(\lambda_i) = \sum_{k=0}^2 c_{jk} x_i^k$, where $x_i$ is a normalized wavelength measure, ranging from -1 to 1 in the fit window. We treat the polynomial coefficients $c_{jk}$ as nuisance parameters and profile over them.

We extract our limits using a Poissonian likelihood for each stacked data set $j$
\beq 
\mathcal{L}_j(\gag, c_{jk}) \propto \prod_i \frac{e^{-\lambda_{ij}}\lambda_{ij}^{n_{ij}}}{n_{ij}!} \enspace,
\label{eq:like_poiss}
\eeq
where the index $i$ runs over wavelength bins in the fit window and $n_{ij}$ are the measured photon counts.
Photon counts in each pixel are treated as independent Poisson variables, as they arise from discrete detection events. The effect of the LSF enters through $\lambda_{ij}$, which we compute by convolving the theoretical ALP signal with the FOS energy response~\cite{XSPECStats2025, Vertongen:2011mu}. 

We combine different stacks by multiplying their respective likelihoods
\beq
 \mathcal{L}(\gag)=\prod_j \mathcal{L}_j(\gag, \hat{\hat{c}}_{jk}(\gag))\enspace,
\label{eq:likecomb_poiss}
\eeq
where $\hat{\hat{c}}_{jk}(\gag)$ are the profiled values of the polynomial coefficients, i.e. the values that maximize $\mathcal{L}_j$ at fixed $\gag$.
We set bounds using the test statistics~\cite{Cowan}
\beq
q(\gag)=
\begin{cases}
-2\ln\!\left(\dfrac{\mathcal{L}(\gag)}{\mathcal{L}(\hat{g}_{a\gamma})}\right)
& \qquad\gag \ge \hat{g}_{a\gamma}\\[6pt]
0
& \qquad\gag < \hat{g}_{a\gamma}\enspace,
\end{cases}\label{eq:q}
\eeq
where $\hat{g}_{a\gamma}$ is the value of the coupling that maximizes the likelihood at fixed ALP mass under the physical constraint $\gag \ge 0$. 
We set 95\% C.L. upper bounds by finding the value of the coupling $\gag$ that corresponds to $q(\gag)=\Delta$. Since some of our spectra contain very low photon counts, we calibrate the threshold $\Delta=2.93$ through the Monte Carlo simulations described in Appendix~\ref{app:monte_carlo}. The bound is slightly more conservative than what one would get under the assumption that $q(\gag)$ follows the asymptotic half $\chi^2$ distribution with one degree of freedom, which would imply a threshold $\Delta=2.71$~\cite{Cowan}.

The 95\% C.L. upper bound derived from the HST FOS data is shown in Figure~\ref{fig:result} by the dark blue band. 
The average value of the bound in the mass range 12.4 to 14.5~eV is $4.6 \times 10^{-12}~\mathrm{GeV}^{-1}$.
The three DM profiles considered yield bounds within 5\% of each other,  with the  \texttt{MilkyWayPotential2022} NFW being the most optimistic and the best-fit NFW from~\cite{2020MNRAS.494.4291C} being the most conservative. 

We show the results of our extended analysis with a light blue band. We find that the model from~\cite{Regis:2024znx} is the most optimistic, while that from~\cite{Roy:2023omw} is the most conservative. These models yield bounds within 14\% of each other.
The Einasto profile from~\cite{2024MNRAS.528..693O} produces bounds that are on average less than 3\% more conservative than \texttt{MilkyWayPotential2022}, suggesting that our bounds are dominated by the DM density at intermediate Galactic radii $5~\mathrm{kpc}<r<15~\mathrm{kpc}$, where the two profiles are similar.
Using an alternative Galactic extinction law from~\cite{1994ApJ...422..158O} improves our bound by a factor of 2.2\%, while increasing the order of the polynomial modeling the background from 2 to 3 leads to a relaxation of the bounds of 6.9\% when averaging over the whole spectrum, but of  only 0.35\% in the interesting region $m_a=12.4-14.5$~eV.

\subsection{IUE}
The IUE low-dispersion spectra are not available as photon counts, since IUE employed SEC vidicon television camera tubes as detectors. As such, IUE records an integrated video image of the spectrum and then reads it out by scanning, rather than counting individual photons.
We then extract our limits using a Gaussian likelihood for each data set
\beq 
\mathcal{L}_j = e^{-\chi^2/2} \;\;\; 
{\rm with} \;\;\; 
\chi^2_j = \sum_{i=1}^{N_{bin}}\left(\frac{S_{th}^i-S_{obs}^i}{\sigma_i}\right)^2\;,
\label{eq:like}
\eeq
where $i$ runs over the wavelength bins, $S_{th}^i$ is the theoretical estimate for the flux density, $S_{obs}^i$ is the observed flux density (minus the continuum emission determined by the arPLS algorithm for IUE data), and $\sigma_i$ is the statistical error. With Eq.~\eqref{eq:like}, we model the per-bin uncertainties with a Gaussian likelihood, with $\sigma_i$ taken from the INES noise model~\cite{SchartelRodriguezPascual1998INESNoiseModel}. Given the long exposure times, each spectral bin is an average over many contributions to the extracted signal, so a Gaussian approximation is adequate for our purposes.

We combine different data sets, labeled by $j$, by multiplying the respective likelihoods
\beq
 \mathcal{L}(\gag)=\prod_j \mathcal{L}_j(\gag)\quad{\rm i.e.,}\quad \chi^2(\gag)=\sum_j \chi^2_j(\gag)\;.
\label{eq:likecomb}
\eeq
We set bounds using the test statistics defined in Eq.~\eqref{eq:q}. We assume $q(\gag)$ follows a half $\chi^2$ distribution with one degree of freedom~\cite{Cowan}, and set  95\% C.L. bounds on $\gag$, by requiring $q(\gag)=2.71$.

Our 95\% C.L. upper bound from the fiducial analysis is shown as a red line in Figure~\ref{fig:result}. The average bound with SWP spectra in the mass range 12.4 to 14.5\,eV is $\gag < 2.3 \times 10^{-12}~\mathrm{GeV}^{-1}$. The orange band indicates the systematic uncertainty due to the modeling of Virgo's DM halo. 
The lower boundary of the band is obtained assuming the most optimistic dark matter profile (largest $\rho_s$ and smallest $r_s$) from~\cite{McLaughlin:1998sb} and that the center of the halo coincides with M87. The upper boundary is obtained with the most pessimistic profile from~\cite{McLaughlin:1998sb} and assuming the center of the halo is $0.5'$ away from our reference center in the direction opposite to M87.

Averaging over ALP masses, the bounds fluctuate by +8.5\% and -11\% compared to the fiducial analysis, with the halo profile and center position contributing variations of +4.1\% and -5.0\%, and 4.3\% and \textcolor{red}{-6.5\%}, respectively, when considered separately. We find that changing the distance to the center of the halo does not affect the bound in any significant way. 
We further test how our bounds change when we use a constant velocity dispersion of 640~km/s and find a modest weakening of the bounds by 2\%. Considering a different Galactic extinction law from~~\cite{1994ApJ...422..158O} worsens the bounds by less than 0.5\%.

Finally, another source of systematic uncertainty is the continuum model. We check how setting the smoothness parameter $\lambda$ of the arPLS algorithm to $10^3$ or $10^7$ affects our bounds. Either choice can strengthen or loosen the bound for a given ALP mass. On average the bounds fluctuate by +1.6\% and -5.0\%  for short-wavelength spectra, and by +5.2\% and -5.0\% for long-wavelength spectra, compared to the fiducial case $\lambda=10^5$.

\section{\label{sec:compare} Comparison with other work}
\begin{figure}[t!]
\centering
   \includegraphics[width=0.7\textwidth]{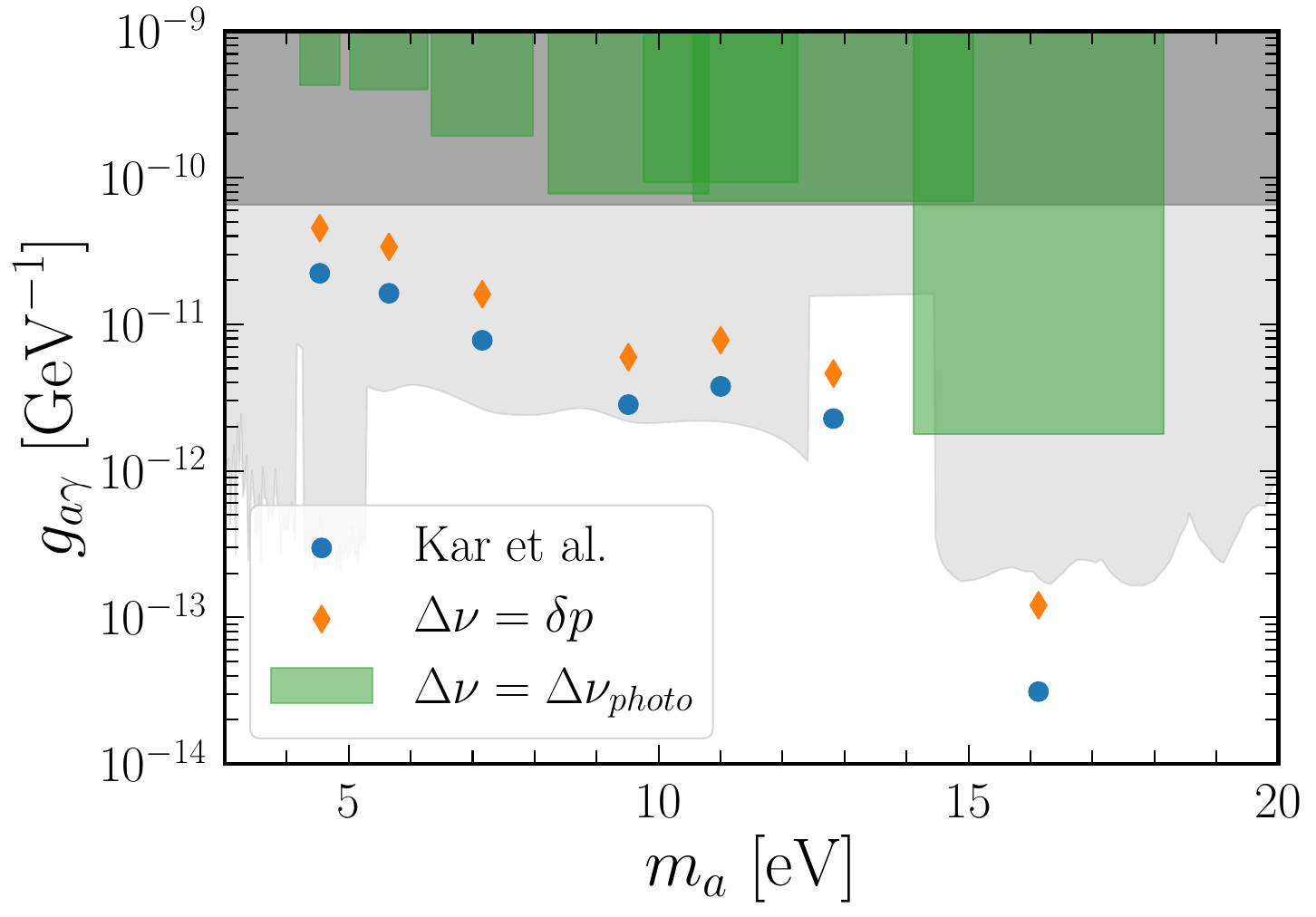}
    \caption{Constraints on the ALP–photon coupling $\gag$ as a function of ALP mass from Swift-UVOT and AstroSat-UVIT observations of M87 (green bands).
Blue dots reproduce the bounds of~\cite{Kar:2025ykb}, while orange diamonds are obtained correcting for aperture interpretation and flux density definition.
Green bands show the correct bounds when accounting for the finite photometric bandwidth, which significantly weakens sensitivity to the narrow spectral line expected from ALP decay. The width of the green bands represents the width of  Swift-UVOT and AstroSat-UVIT photometric bands.
    }
\label{fig:bound_swift_astrosat}
 \end{figure}

Reference~\cite{Kar:2025ykb} reports strong bounds in an ALP mass range overlapping with ours, based on data from~\cite{EventHorizonTelescope:2024uoo}. Here we revisit those constraints.

In April 2018, a coordinated multi-wavelength campaign involving over two dozen observatories, including the Event Horizon Telescope (EHT), observed the supermassive black hole M87*, capturing data from radio to very high-energy gamma-rays.
The relevant data points for us are those acquired with Swift-UVOT (6 data points) and AstroSat-UVIT (1 data point). These data are listed in Table 1 of~\cite{Kar:2025ykb}, which is in turn taken from Table C.1 of~\cite{EventHorizonTelescope:2024uoo}.
Swift-UVOT observed M87 in its optical ($v$, $b$ and $u$) and UV ($w1$, $m2$ and $w2$) photometric
bands. The central frequencies and bandwidths of Swift-UVOT can be found in Table 1 of Ref.~\cite{Poole:2007xi}. The resolution $\Delta\lambda / \lambda$ ranges from 0.14 to 0.34, much larger than the DM velocity dispersion. The AstroSat-UVIT datapoint is obtained with the BaF2 filter, with a  $\Delta\lambda / \lambda$  of about 0.25.

In the spirit of comparing with~\cite{Kar:2025ykb}, we adopt the same dark matter profile. We assume an NFW profile for the Virgo dark matter halo with $\rho_s = 6.96 \times 10^5~M_\odot/\mathrm{kpc}^{-3}$, $r_s = 403.8~\mathrm{kpc}$, and a distance of 17.2~Mpc~\cite{HAWC:2023bti}. We take the center of the halo to coincide with the position of M87* and assume that the observations are centered there. We remove a region of 100 Schwarzschild radii around the BH.

First, we reproduce the results of~\cite{Kar:2025ykb}. To calculate the flux from ALP decay, we use their Eq. (2.3) 
\beq
\langle S_\nu \rangle 
= \Gamma_a \int d\ell\,d\Omega\, B(\Omega)\,\frac{\rho_a}{\delta p}
 = \frac{\Gamma_a}{\langle\delta p\rangle} \tilde{D} 
\label{eq:Snu_kar}
\enspace.
\eeq
Here, the velocity dispersion $\delta v = \delta p / m_a$ has two components: a halo component set to $10^{-3}$, and a black hole component as described in Section~\ref{sec:data}.

With this setup, we are able to accurately reproduce the bounds obtained with the ``conservative approach" of~\cite{Kar:2025ykb}. This approach consists in requiring that the flux density from ALP decay does not exceed the measured one. We match their results by assuming: 1) top-hat circular apertures with a radius of $3''$ for Swift-UVOT and of $300''$ for AstroSat-UVIT, 2) measured flux density obtained by dividing $\nu S_\nu$ reported in Table 1 of~\cite{Kar:2025ykb} (or Table C.1 of~\cite{EventHorizonTelescope:2024uoo}) by the central frequency of the respective frequency band.

From Table C.1 of~\cite{EventHorizonTelescope:2024uoo}, we see however that $3''$ and $300''$ are the FWHM angular extents over which the flux is collected. If these values are interpreted as diameters, rather than radii, of the apertures, we obtain the modified D-factors
$\tilde{D} = 6.2 \times 10^{22}~\mathrm{eV / cm}^2$ 
for the $3''$ aperture and 
$\tilde{D} = 3.7 \times 10^{26}~\mathrm{eV / cm}^2$ 
for the $300''$ one. These modified D-factors are a factor of 3.9 and 3.3 smaller than those resulting from taking $3''$ and $300''$ as radii.
Additionally, for the AstroSat-UVIT observation, the flux density in the sixth column of Table C.1 of~\cite{EventHorizonTelescope:2024uoo} corresponds not to $\nu S_\nu$ divided by the central frequency, but by the bandwidth. Taking into account the correct $\tilde{D}$ and taking the observed flux density from the sixth column of Table C.1 of~\cite{EventHorizonTelescope:2024uoo}, we obtain the bounds shown in Fig.~\ref{fig:bound_swift_astrosat} as orange diamonds. These are a factor of 2 and 4 weaker for Swift and AstroSat, respectively, than reported in~\cite{Kar:2025ykb}.

We now address the central issue: Eq.~\eqref{eq:Snu_kar} is not the correct expression to compute the flux in a photometric band that is significantly broader than the intrinsic bandwidth of the ALP signal. Instead, one should use Eq.~\eqref{eq:Snu_large_bw}\footnote{We do not include extinction or the frequency dependence of the beam.}, with $\Delta\nu = \Delta\nu_{photo}$, the bandwidth of the photometric filter, because this is the frequency range over which the ALP signal gets averaged.
The bounds obtained this way are shown in Fig~\ref{fig:bound_swift_astrosat} as green bands. They are weaker by factors of 10–15 compared to the orange diamonds, leading to a total weakening of a factor of 20–30 for Swift and about 60 for AstroSat, relative to the original bounds of~\cite{Kar:2025ykb}.
Additional weakening may come from assuming a different center and profile of the dark halo, as discussed in Section~\ref{sec:IUE}.

Lastly, we comment on the bound obtained in~\cite{Kar:2025ykb} from ``historical data" consisting of six data points.
In the UV region, these bounds are obtained from IUE data, presumably from rebinning the spectra used in Ref.~\cite{1980ApJ...240..447P}. That study focused on emission from the nucleus and jet of M87 using four IUE spectra—specifically, swp03571, swp04299, lwr3159, and lwr03818—which are also part of the data set we analyzed in Section~\ref{sec:IUE}. Our bounds are based on these and additional observations analyzed at full spectral resolution. As a result, our limits are likely to be more stringent than what can be obtained from rebinning a smaller set of the data.

\section{\label{sec:conc} Conclusions}
In this work, we searched for a spectral line from the decay of axion-like particle (ALP) dark matter in ultraviolet data from the Hubble Space Telescope (HST) and the International Ultraviolet Explorer (IUE). We focused on the 12.4–14.5~eV ALP mass range, which remains relatively unexplored and is within the sensitivity window of these instruments.

Using archival blank-sky data from the HST Faint Object Spectrograph (FOS) and observations of M87 from IUE, we derived new upper bounds on the ALP–photon coupling $\gag$. Our most stringent limit, obtained from IUE, excludes values of $\gag$ above $2.3 \times 10^{-12}~\mathrm{GeV}^{-1}$, or ALP lifetimes below $10^{25}$~s, and improves upon previous constraints by a factor of 7. HST blank sky observations provide a complementary bound at the level of $4.6 \times 10^{-12}~\mathrm{GeV}^{-1}$. Although the HST FOS had better sensitivity than the IUE spectrographs, the bounds obtained with the two instruments are comparable due to the lower D-factor of the Milky Way compared to Virgo and to the poorer spectral resolution of the low-dispersion configurations of the FOS. The narrower instrumental response of IUE better matches the intrinsic width of the ALP decay line, leading to a more favorable signal-to-noise ratio despite lower sensitivity.

We also revisited recent claims in the literature based on photometric data from Swift-UVOT and AstroSat-UVIT. We showed that these bounds are significantly overestimated due to incorrect treatment of the ALP spectral line width relative to the broad instrumental bands. When properly accounting for spectral resolution, we find the resulting constraints are weakened by well over one order of magnitude.

Our analysis demonstrates the power of spectroscopic data for probing ALP dark matter in the extreme ultraviolet and highlights the importance of accurate spectral modeling. Future UV observatories with improved sensitivity may further extend the reach of this method into new regions of parameter space.

\appendix
\section{FOS data reduction and forward modeling of the ALP signal}\label{app:data_reduction}

\subsection{Data reduction}
\begin{figure}[t!]
\centering
   \includegraphics[width=1\textwidth]{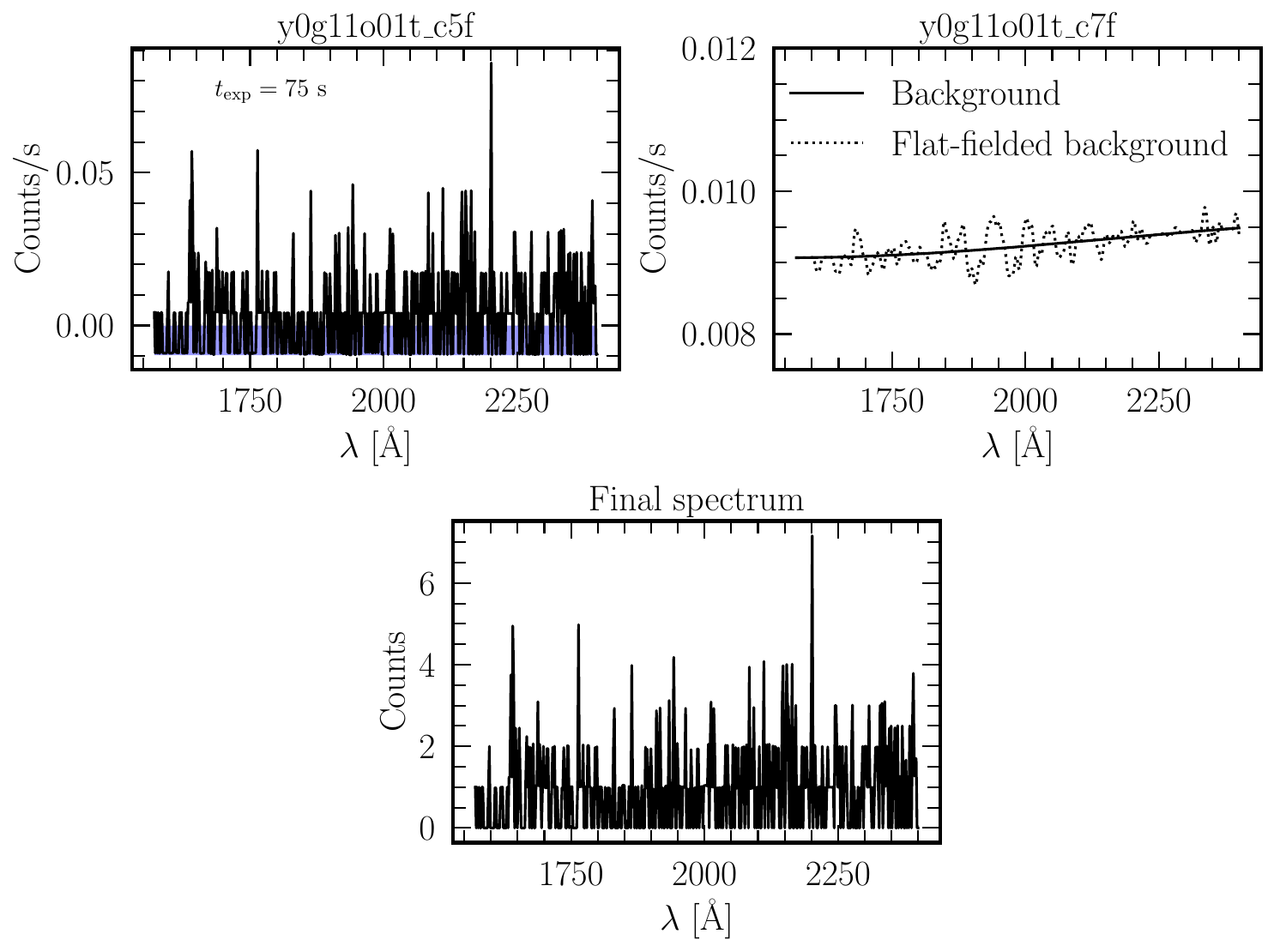}
    \caption{Construction of the spectra analyzed in this work from intermediate products of the FOS calibration pipeline. The flat-fielded, background-subtracted spectrum \texttt{c5f} and the flat-fielded dark background are added and multiplied by the per-pixel exposure to obtain the final spectrum. We take the dark background from the \texttt{c7f} file and multiply it by the flat field response, stored in a calibration reference file, to ensure consistency with \texttt{c5f}. Negative counts seen in \texttt{c5f}, highlighted in the top-left panel by a blue shade, are not present in the final spectrum.}
\label{fig:spectrum_construction_FOS}
 \end{figure}

The FOS data calibration pipeline (\texttt{calfos}) is structured as follows~\cite{FOSHandbook1998}. The raw data, consisting of photon counts per pixel, is read into memory, errors and data quality are initialized, and the spectrum is converted into count rates. The error in each pixel is taken to be simply the square root of the number of detected photons. These errors underestimate the actual statistical uncertainty if the photon counts are low and we do not rely on them in our analysis. 

After shifting the spectrum to compensate for geomagnetically-induced image motion and correcting for detector saturation, the dark counts (photon counts that would be detected even if no light of astrophysical origin were reaching the detector) are subtracted. The primary source of the dark photon counts is Cherenkov radiation generated by high-energy cosmic rays and electrons from the Earth's radiation belts as they pass through the faceplate covering the photocathode. Since for most observations a dedicated dark measurement was not performed, the background is modeled using a reference file known as the \texttt{BACHFILE}, which contains a baseline dark count rate~\cite{Lyons1992FOS080}. 
Because the intensity of this particle flux varies with the telescope's geomagnetic position, the dark count rate from the \texttt{BACHFILE} is scaled by an orbit-dependent factor, derived from the spacecraft’s geomagnetic coordinates at the time of observation. 
The typical dark rate for the red detector is 0.01 counts/s/pixel and shows little variation over the diode array (see Figure~\ref{fig:spectrum_construction_FOS}). The FOS data handbook~\cite{FOSHandbook1998} reports that the dark background might be underestimated in the \texttt{BACHFILE} by up to 30\%.  At the end of this calibration step, the subtracted background is saved to a file (\texttt{c7f} or \texttt{c7h}).

Next, the background-subtracted spectrum is multiplied by the flat-field response, which corrects for sensitivity variation from diode to diode and is stored in the \texttt{FL1HFILE}.  The resulting spectrum is saved to file (\texttt{c5f} or \texttt{c5h}). The pipeline also has tasks for subtraction of scattered light and continuum sky emission. However, these two tasks were not performed on the data considered here.
The pipeline ends with wavelength and absolute flux calibration, the latter of which we describe below.

The spectra used in this work are obtained by multiplying the subtracted dark background (\texttt{c7f}) by the flat field response, adding it back to the \texttt{c5f} file, and multiplying the result by the per-pixel exposure to obtain a spectrum in units of counts. This guarantees that we have no wavelength bins with negative counts, which are generally present after background subtraction, especially for short exposures. In our statistical analysis, we model the dark background, along with any continuum sky emission, as a second-order polynomial. An example of our construction is shown in Figure~\ref{fig:spectrum_construction_FOS}.
Finally, we stack spectra with the same line of sight by summing the photon counts in each pixel.

\subsection{Forward modeling}

To construct our forward model operator, i.e., to convert the ALP-decay flux into detector counts, we invert the last two steps of the FOS calibration pipeline, namely, the absolute flux calibration (\texttt{AIS\_CORR}), and the time-dependent sensitivity correction (\texttt{TIM\_CORR}). When sky subtraction is omitted, as for our data, these steps, along with the aperture throughput and focus corrections (\texttt{APR\_CORR}), convert the \texttt{c5f} count-rate spectrum into the final flux-calibrated product \texttt{c1f}.

In \texttt{calfos}, the \texttt{APR\_CORR} step is applied before \texttt{AIS\_CORR}. \texttt{APR\_CORR} rescales data to match what would be seen through the reference \texttt{4.3} aperture and applies a focus-dependent throughput correction derived from the telescope focus history. These corrections are intended to place observations taken with different apertures and focus states onto a common photometric system so that a single average inverse sensitivity curve can be used in the subsequent \texttt{AIS\_CORR} step.

In our forward operator, we omit the inversion of \texttt{APR\_CORR} for physical reasons. Firstly, our data was taken with the reference aperture. Secondly, our expected signal is uniform over the solid angle subtended by the aperture. For a point source, changes in telescope focus modify the point-spread function (PSF) and therefore the fraction of light transmitted through the finite entrance aperture, producing an apparent sensitivity variation that \texttt{APR\_CORR} is designed to correct. In contrast, for a uniform extended source that fills the aperture, the recorded signal is proportional to the surface brightness integrated over the aperture solid angle and is insensitive to PSF redistribution. Light blurred out of the aperture from one part of the source is compensated for by light from adjacent regions being blurred into the area. 

\texttt{AIS\_CORR} then converts \texttt{c5f} spectra, in units of counts/s, into absolute flux units (erg/s/cm$^2$/\AA) by multiplying the counts by the point-source inverse sensitivity $\rm IVS_{p}(\lambda)$. The inverse sensitivity is an average over data from all calibration observation epochs and is stored in a reference file called \texttt{AISHFILE}. The FOS sensitivity curve contains an aperture throughput factor $T_{4.3}(\lambda)$ which accounts for the portion of the PSF that falls outside the standard $4.3''$ reference aperture.
The $T_{4.3}$ is introduced so that the calibrated flux reflects the total flux of a point source. For extended sources that uniformly fill the aperture, like the emission from DM decay in our galaxy, this term is not necessary (see Ref.\cite{FOSHandbook1998} Section~32.6.6). We thus multiply the inverse sensitivity $\rm IVS_{p}(\lambda)$ from the \texttt{AISHFILE} by $T_{4.3}$ to remove this factor and recover the extended sensitivity $\rm IVS_e(\lambda)$. 
We obtain $T_{4.3}(\lambda)$ from the pre-COSTAR \texttt{synphot}~\cite{Synphot2005} throughput reference file \texttt{fos\_sqr4p3\_005\_syn}, available from the HST Calibration Reference Data System (CRDS)~\cite{Greenfield2016CRDS}. At the wavelengths considered in our work, this wavelength-dependent factor is of order 0.55.

During the final step, the \texttt{TIM\_CORR} task corrects for time-dependent drifts in detector sensitivity by dividing the data by a correction factor $T(\lambda)$ derived from the \texttt{CCSD} reference file.

In summary, the flat-fielded, background-subtracted spectrum c5f relates to the surface brightness of a diffuse source as follows
\beq
B(\lambda) = c(\lambda)\frac{ {\rm IVS_e(\lambda)}}{T(\lambda)\Omega}\enspace,
\eeq
where $c(\lambda)$ is \texttt{c5f} count rate, and $\Omega$ is the projected solid angle of the aperture ($4.3'' \times 1.43''$ in our case).
Conversely, to turn the predicted ALP decay flux collected in the aperture into photon counts at the detector, we use the forward operator $R(\lambda)$, defined in Eq.~\eqref{eq:forward}. We explicitly check on our data that multiplying the \texttt{c5f} spectrum by ${\rm IVS_p}(\lambda)/(T(\lambda)F(\lambda))$ yields the final flux-calibrated spectrum from the pipeline (\texttt{c1f}), where $F(\lambda)$ is the focus correction from \texttt{APR\_CORR}. To maintain consistency when combining multiple FOS data sets into a single stacked count spectrum, we construct the total stack forward operator by summing the individual $R(\lambda)$ components of each constituent exposure.

\section{Monte Carlo tests of asymptotic statistics for FOS data}\label{app:monte_carlo}

\begin{figure}[t!]
\centering
   \includegraphics[width=0.8\textwidth]{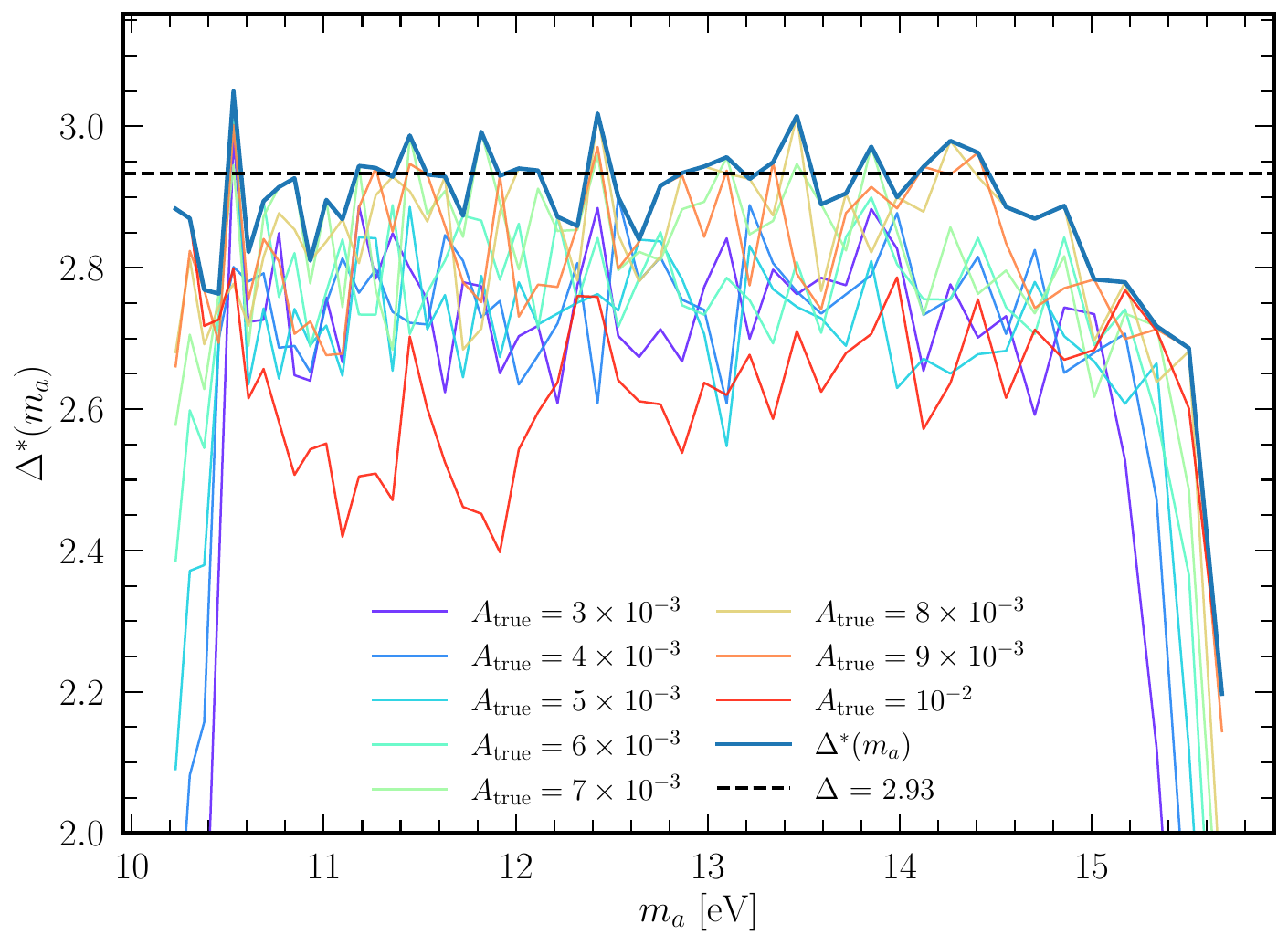}
\caption{
Calibration of the one-sided 95\% C.L.\ threshold $\Delta$ used to set upper limits.
For each ALP mass $m_a$, we generate Monte Carlo realizations and determine the minimum threshold $\Delta^*(m_a;A_{\rm true})$ required to achieve $\ge 95\%$ coverage for each injected signal amplitude $A_{\rm true}$ (thin colored curves, $A_{\rm true}=3\times10^{-3}$ to $10^{-2}$).
We then take the median value of $\Delta^*(m_a)$ in the central mass region 11-14.5~eV (thick blue curve).
The horizontal dashed line shows the single global choice $\Delta=2.93$ adopted for the final limits.
}
\label{fig:delta_threshold}
 \end{figure}

In the main analysis, we set one-sided 95\% C.L. upper limits by solving for the coupling at which the test statistic $q(\gag)$ increases by a fixed threshold $\Delta=2.93$ above its best-fit value. Here we provide details on how we determine this threshold.

For sufficiently regular likelihoods and in the asymptotic limit (large sample), one expects $\Delta=2.71$ for a one-sided 95\% C.L. bound with one parameter of interest~\cite{Cowan}. However, because several of our stacked spectra contain very low photon counts, the asymptotic approximation can be imperfect. We therefore calibrate $\Delta$ using Monte Carlo simulations that reproduce the full analysis chain.

For each tested ALP mass $m_a$, we first construct a truth model for the counts under the background-only hypothesis. Concretely, we fit the background-only model $g_{a\gamma}=0$ following the same procedure as in the main analysis, as described in Section~\ref{sec:res_FOS}. This yields, for each stacked spectrum $j$, a best-fit background expectation $\hat{\lambda}^{(0)}_{ij}$.

\begin{figure}[t!]
\centering
   \includegraphics[width=0.8\textwidth]{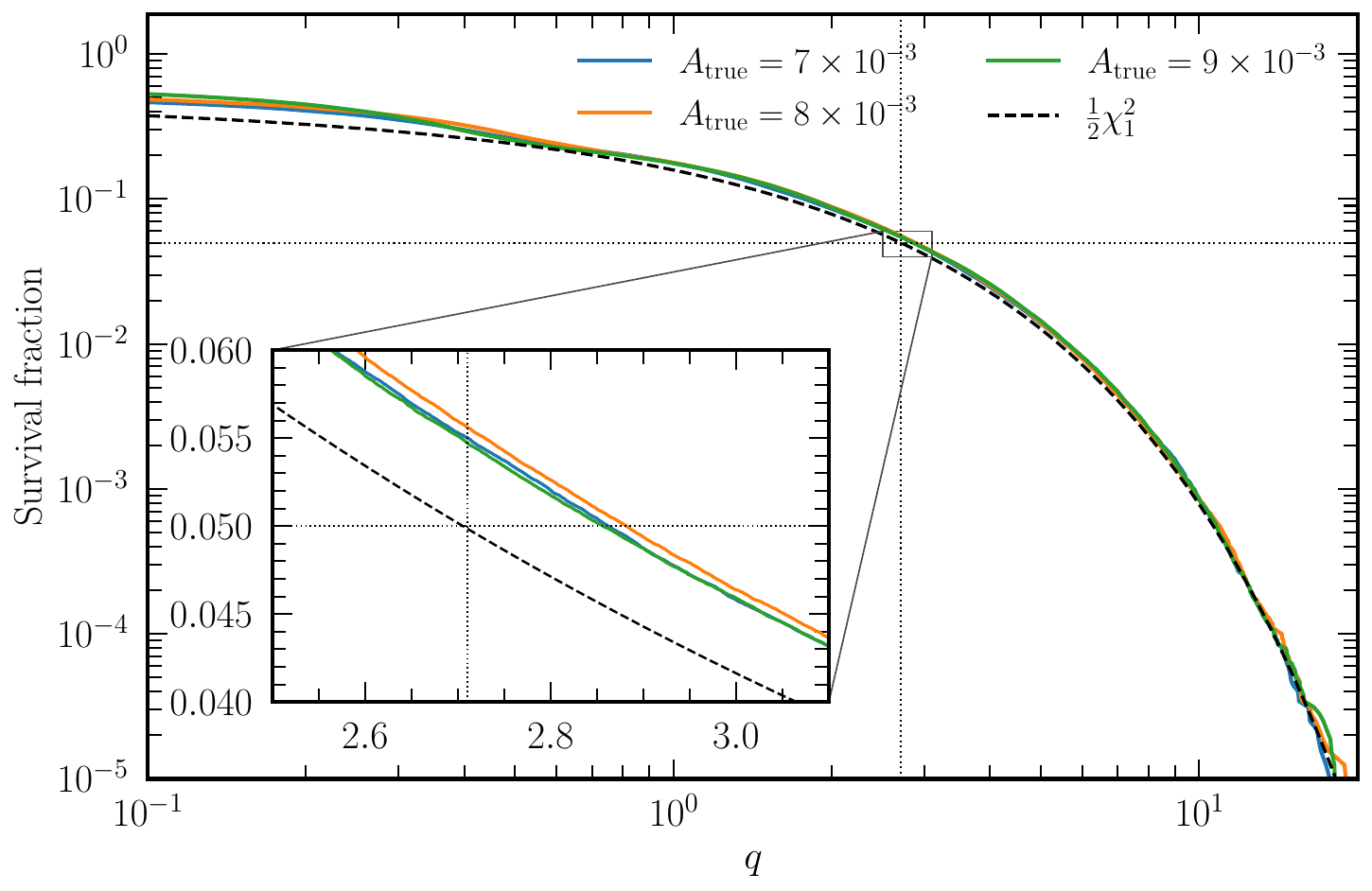}
\caption{
Survival function of the one-sided upper-limit test statistic $q$ from Monte Carlo realizations, shown for the injected amplitudes $A_{\rm true}$ that drive the calibration of the upper-limit threshold $\Delta$.
For each $A_{\rm true}$, the survival fraction is computed by pooling toy data sets over the tested masses on the coarse mass grid that are within $m_a=11$--$14.5~\mathrm{eV}$.
The black dashed curve shows the asymptotic expectation $\tfrac12\chi^2_1$.
The vertical dotted line marks the nominal one-sided 95\% threshold $q=2.71$, and the horizontal dotted line marks a survival fraction of 0.05.
}
\label{fig:survival}
 \end{figure}

We then generate Monte Carlo realizations of the counts by drawing independent Poisson counts in each pixel
$n^{\rm toy}_{ij} \sim {\rm Poisson}(\mu_{ij})$,
where the mean $\mu_{ij}$ is taken to be the background-only expectation plus an optional injected signal
\beq
\mu_{ij} = \hat{\lambda}^{(0)}_{ij} + A_{\rm true}C_{ij}\enspace,
\qquad A_{\rm true}\equiv g_{10,\rm true}^2\ge 0\enspace.
\eeq
Here $C_{ij}$ is the signal template expressed in expected counts for $g_{10}=1$, evaluated exactly as in the main data analysis. We generate toys both for $A_{\rm true}=0$, and for a set of nonzero injections $A_{\rm true}$ chosen to bracket our upper limits. To reduce computational time, we perform the calibration on a coarse mass grid, using one ALP mass every ten points of the analysis grid, resulting in 50 tested masses.  We repeat the procedure for $A_{\rm true}= \{2, 3,4,5,6,7,8,9,10\}\times 10^{-3}$, corresponding to $g_{a\gamma,\rm true}=(4.47-10.0) \times 10^{-12}~\mathrm{GeV}^{-1}$
For each combination of $m_a$ and $A_{\rm true}$, we generate 10,000 toy spectra.

For each toy realization, we run the full analysis procedure of Section~\ref{sec:res_FOS} and compute the profiled test statistics $q$ of Eq.~\eqref{eq:q} as a function of signal amplitude $A=g_{10}^2$ on the same wavelength grid used in the main analysis. From this curve, we obtain the one-sided upper limit $A_{\rm UL}$ by finding the smallest $A\ge \hat{A}$ such that $q=\Delta$, with the physical constraint $\hat{A}\ge 0$ imposed throughout. For a given injected amplitude $A_{\rm true}$, the frequentist coverage probability at threshold $\Delta$ is defined as 
\beq
{\rm Coverage}(\Delta) \equiv P(A_{\rm UL}(\Delta) \ge A_{\rm true})\enspace,
\eeq
estimated by the fraction of toys for which the upper limit actually exceeds the injected value.

To calibrate $\Delta$, for each tested combination of $m_a$ and $A_{\rm true}$, we determine the smallest value of $\Delta$ that yields at least 95\% coverage. We then adopt, for that mass, the most conservative (largest) of these thresholds across all injected amplitudes, ensuring that the final $\Delta$ controls the worst-case undercoverage among the tested injections.

The procedure above yields a set of threshold values $\Delta^*(m_a)$ shown in Figure~\ref{fig:delta_threshold}. In practice, we adopt a single mass-independent threshold for the final limits. We choose the global $\Delta$ threshold as the median of $\Delta^*(m_a)$ in the central mass range 11-14.5~eV. The resulting global $\Delta$ is then used in the main pipeline to compute the reported upper limits at all masses. For 41\% of the tested masses within 11--14.5~eV, the value of $\Delta$ is determined by toy models with $A_{\rm true}= 8\times 10^{-3}$ ($g_{a\gamma,\rm true}=8.9\times 10^{-12}~\mathrm{GeV}^{-1}$), while $A_{\rm true}= 7\times 10^{-3}$ and $A_{\rm true}= 9\times 10^{-3}$ account for the value of $\Delta$ for another 47\% of the cases. The survival fraction, pooled over tested ALP masses in the central mass range, is shown in Figure~\ref{fig:survival}.

\section*{Acknowledgements}
ET is grateful to Marco Regis, Anne Green, Alfonso Aragon-Salamanca, Marco Taoso, and Nicolao Fornengo for useful discussions.
ET is supported by STFC Consolidated Grant [ST/T000732/1]. This article/publication is based upon work from COST Action COSMIC WISPers CA21106, supported by COST (European Cooperation in Science and Technology). 

\textbf{Data Availability Statement}: The data used in this work is publicly available from the Mikulski Archive for Space Telescopes (MAST)~\cite{mast} or the IUE Newly Extracted Spectra (INES) system~\cite{ines}. The stacked FOS count spectra are available at \href{https://github.com/elisabm99/UV-legacy}{https://github.com/elisabm99/UV-legacy}.

\bibliographystyle{JHEP_improved}
\bibliography{biblio}

\end{document}